\documentclass[prd,twocolumn,nofootinbib,showpacs,superscriptaddress]{revtex4-1}

\usepackage{amsfonts}
\usepackage{amsmath}
\usepackage{amssymb}
\usepackage{bm}
\usepackage{dcolumn}
\usepackage{graphicx}
\usepackage{graphics}
\usepackage[latin1]{inputenc}
\usepackage{latexsym}
\usepackage{rotating}
\usepackage{xspace} % Sensible space treatment at end of simple macros
\usepackage[usenames,dvipsnames]{color}
\usepackage{mathrsfs}
\usepackage{subfigure}
\usepackage{yfonts}
\usepackage{ulem}
\usepackage{outlines}
\usepackage{enumitem}
\setenumerate[1]{label=\Roman*.}
\setenumerate[2]{label=\Alph*.}
\setenumerate[3]{label=\roman*.}
\setenumerate[4]{label=\alph*.}

\definecolor {darkgreen}{rgb}{0.2,0.7,0.2}

\newcommand\be{\begin{equation}}
\newcommand\ba{\begin{eqnarray}}
\newcommand\ee{\end{equation}}
\newcommand\ea{\end{eqnarray}}

\newcommand\bw{\begin{widetext}}
\newcommand\ew{\end{widetext}}

\newcommand{\nn}{\nonumber}

\newcommand{\MAT}{{\mbox{\tiny mat}}}

\newcommand{\CS}{{\mbox{\tiny CS}}}

\newcommand{\dcs}{{\mbox{\tiny dCS}}}
\newcommand{\gr}{{\mbox{\tiny GR}}}

\newcommand{\podi}{{\mbox{\tiny pole-dipole}}}
\newcommand{\dd}{{\mbox{\tiny dd}}}
\newcommand{\QUAD}{{\mbox{\tiny Q}}}
\newcommand{\BH}{{\mbox{\tiny BH}}}
\newcommand{\MQ}{{\mbox{\tiny MQ}}}
\newcommand{\ORB}{{\mbox{\tiny orb}}}
\newcommand{\BL}{{\mbox{\tiny BL}}}
\begin{document}
\title{Spin-Precessing Black Hole Binaries in Dynamical Chern-Simons Gravity}

\author{Nicholas Loutrel}
\affiliation{eXtreme Gravity Institute, Department of Physics, Montana State University, Bozeman, MT 59717, USA.}

\author{Takahiro Tanaka}
\address{Department of Physics, Kyoto University, Kyoto 606-8502, Japan}
\affiliation{Yukawa Institue for Theoretical Physics, Kyoto University, 606-8502, Kyoto, Japan}

\author{Nicol\'as Yunes}
\affiliation{eXtreme Gravity Institute, Department of Physics, Montana State University, Bozeman, MT 59717, USA.}

\date{\today}

%%%%%%%%%%%%%%%%%%%%%%%%%%%%%%%%%%%%%%%%%%%%%%%%%
\begin{abstract} 

Gravitational waves from spin-precessing binaries exhibit amplitude oscillations that provide an invaluable method to extract the spins of the inspiraling compact objects. The spin-spin and spin-orbit interactions that cause this effect are sensitive to the fundamental nature of gravity, which will allow us to constrain modified theories of gravity using gravitational wave observations of precessing binaries. We here consider precessing black hole binaries in dynamical Chern-Simons gravity, an effective theory of gravity that enhances parity violating interactions. We model the black holes as modified point particles using effective field theory, and derive the spin-precession equations for a binary system by working within the post-Newtonian formalism. We find that the spin-spin and quadrupole-monopole interactions of General Relativity are modified due to an interaction between the scalar dipoles of the two black holes and the modified black hole quadrupole as a result of the violation of the no hair theorems. These modifications enter the precession equations at leading post-Newtonian order. We further show that these precession equations admit seven constants of motion when neglecting radiation reaction, with only the mass-weighted effective spin being modified from General Relativity. We discuss how these may be used to reduce the precession equations to quadrature and the possibility of constructing analytic Fourier domain waveforms for generic spin-precessing binaries in dCS gravity.

\end{abstract}

%\pacs{04.30.-w,04.50.Kd,04.25.-g,97.60.Jd}

%04.30.Db Wave generation and sources
% 04.50.Kd Modified theories of gravity
% 04.25.-g Approximation methods; equations of motion
%04.25.Nx Post-Newtonian approximation; perturbation theory; related approximations
%97.60.Jd Neutron stars

\maketitle

%%%%%%%%%%%%%%%%%%%%%%%%%%%%%%%%%%%
\section{Introduction}
\label{intro}

The detections of gravitational waves (GWs) by the Advanced Laser Interferometer Gravitational Wave Observatory (aLIGO)~\cite{Abramovici:1992ah, Abbott:2007kv, Harry:2010zz, ligo} and the Advanced Virgo detectors~\cite{Caron:1997hu, Giazotto:1988gw, TheVirgo:2014hva, virgo} has ushered in a new era of astrophysics and precision tests of Einstein's theory of General Relativity (GR). Through these detections, we have uncovered black holes (BHs) with masses higher than those observed before hand~\cite{GW150914}, more precisely pinned down the event rates for the coalescence of compact objects, placed significant constraints on parameterized post-Einsteinian (ppE) parameters describing generic deviation from GR~\cite{PPE,Yunes:2016jcc}, provided some of the earliest constraints on the equation of state of dense nuclear matter in the cores of neutron stars (NSs)~\cite{Annala:2017llu, Radice:2017lry}, and confirmed the hypothesis that kilonovae are associated with binary NS mergers~\cite{TheLIGOScientific:2017qsa, Tanvir:2017pws}. Yet, there is still much that GWs may teach us.

In the case of BH binaries, the spin angular momentum can play an important role in the dynamics, and subsequent GW emission, of the system. Supernovae, in which these BHs are created, introduce torques on the system, resulting in a misalignment between the spin of the remnant and the orbital angular momentum~\cite{Spruit:1998sg, Wang:2005jg, Wang:2006zia, Kalogera:1999tq, Gerosa:2014kta, Kaplan:2008qm}. For a co-evolving binary, two such events may occur, and there are strong arguments to expect misalignment between spin and orbital angular momentum. Gravitational interactions couple the spins of the BHs to the orbital angular momentum, hereafter referred to as \textit{spin-orbit interactions}, and to each other, \textit{spin-spin interactions}. In the event of misalignment, these interactions cause the spins and orbital angular momentum to precess and nutate about an axis defined by the total angular momentum~\cite{Thorne:1984mz, Hawking:1987en, PW}. Such precession is important because the GWs emitted by the system are beamed along the direction of the orbital angular momentum. 

For precessing binaries, the precession of the orbital angular momentum causes the amplitude of the observed GWs to modulate with a period much longer than the orbital period of the binary, but shorter than its inspiral time~\cite{Apostolatos:1994ha, Apostolatos:1994mx}. Such an effect becomes important for strongly spinning systems that are observed over thousands of orbital cycles, allowing one to capture many of the precession cycles. While arguably more important for low frequency GW detectors such as the Laser Interferometer Space Antenna (LISA)~\cite{Danzmann:2003tv} or the Deci-hertz Interferometer Gravitational wave Observatory (DECIGO)~\cite{Kawamura:2011zz}, these effects are still strong enough to be a required part of templates for searches with ground-based detectors, and have been found to break degeneracies between mass and spin, allowing us to distinguish between BHs, NSs, and other exotic compact objects~\cite{Chatziioannou:2014coa}.

The modulation of the GW amplitude depends sensitively on the nature of the spin-orbit and spin-spin interactions, which in turn, depends on the theory of gravity one considers. Therefore, the amplitude modulation may be different from GR if gravity is fundamentally different from Einstein's prediction. Such an effect becomes particularly important in theories of gravity where GR deviations are degenerate with the magnitude of the spin angular momentum of the binary components. In fact, such degeneracies can be so strong that they have prevented ground-based instrument from being able to place stringent constraints on some modified gravity alternatives~\cite{Yunes:2016jcc,Alexander:2017jmt}. Modifications to the amplitude modulation of GWs in spin-precessing binaries, however, can break this degeneracy, and thus, become a powerful probe of the fundamental nature of gravity.

One example of a modified theory of gravity where this is the case is dynamical Chern-Simons (dCS) gravity~\cite{jackiw, Alexander:2009tp}, an effective field theory that is motivated from string theory~\cite{Polchinski:1998rr}, quantum gravity~\cite{Alexander:2004xd, Taveras:2008yf, Calcagni:2009xz}, and inflation~\cite{Weinberg:2008hq}. DCS gravity modifies the Einstein-Hilbert action through a dynamical pseudo-scalar field which couples to the Pontryagin density, a quadratic curvature invariant constructed from the Riemann tensor and its dual, thus making it a subset of the much larger set of theories referred to as quadratic gravity. This theory is said to be parity violating, in the sense that modifications to GR only appear in systems that are odd under parity transformations, and thus have a preferred axis. One such system is a binary composed of spinning BHs, which has been well studied in this theory, with isolated spinning BH solutions known to fifth order in a small spin expansion~\cite{yunespretorius,kent-CSBH,Maselli:2017kic} and in the near-extremal limit~\cite{McNees:2015srl}.

Placing constraints on the coupling constants of dCS gravity has proven to be difficult, due to modifications only appearing in parity-odd systems~\cite{Yunes:2008ua, Alexander:2007zg, Smith:2007jm, Alexander:2007vt, Yagi:2013mbt}. Currently the most stringent constraints come from measurements of the frame-dragging effect around Earth with the LAGEOS and Gravity Probe B missions. These constraints limit the (dimensional) coupling constant $\xi^{1/4} \lesssim 10^{8} \; {\rm km}$, one of the weakest for a modified theory of gravity that passes Solar System tests~\cite{alihaimoud-chen}. Thus, tests of GR in the dynamical strong field regime of gravity, which is probed by the emission of GWs, appear to be our best bet for constraining dCS gravity. Investigations of extreme- and intermediate-mass ratio inspirals revealed modifications to the balance laws describing the loss of orbital energy due to GWs~\cite{Panietal, prisgair, Sopuerta:2009iy}, as a result of the emission of dipole radiation not present in GR. Quasi-circular inspirals of comparable mass BH binaries have been studied in~\cite{Yagi:2012vf}, with the far-zone metric perturbation computed in~\cite{Yagi:2011xp}. The axial modes of non-rotating BHs become modified in dCS gravity, producing observable effects within the ringdown phase of a binary coalescence~\cite{Yunes:2007ss, motohashi2, cardosogualtieri}. 

However, all of these studies have assumed non-precessing binaries, where the spins of the BHs are either aligned or anti-aligned with the orbital angular momentum. If we relax this assumption, then the spin-orbit and spin-spin interaction is modified from those in GR, introducing new observable features in the waveform. Thus, the inspiral of spin-precessing binaries may provide some of the strongest constraints yet on dCS gravity. The goal of this paper is to derive the modified precession equations necessary to calculate the waveforms for spin-precessing binaries in dCS gravity.

To model the conservative dynamics of spinning BH binaries in dCS gravity, we modify the \textit{effective field theory} (EFT) method of GR used to model the motion of compact objects. The foundations of modern EFT were first laid out by Mathisson~\cite{Mathisson2010} through the so-called \textit{gravitational skeleton}~\cite{Tulczyjew:1957, Tulczyjew:1959, ehlers1979isolated}, a multipole-style expansion of the stress energy tensor motivated by similar expansions of charge density in electromagnetism~\cite{Steinhoff:2010zz}. 

Modern EFTs for the modeling of gravitating systems have a hierarchical structure, due to the separation of multiple scales throughout the problem, specifically the $R_{\rm s} \ll R_{\rm orb} \ll \lambda$, where $R_{\rm s}$ is the characteristic scale of the compact objects, $R_{\rm orb} $ is the orbital radius, and $\lambda$ is the wavelength of gravitational radiation, the observable emission from the system~\cite{Goldberger:2007hy}. At scales $R_{\rm s}$ much smaller than the scale of the wave zone, one can integrate out the internal structure of compact objects, constructing a point particle EFT that describes the degrees of freedom of the objects. At the wave zone scale $\lambda$, it is also convenient to integrate out short range interactions between the compact objects, producing a composite systems described by a set of multipole moments which source the GW emission. In this work, EFT will specifically refer to the point particle EFT of compact objects.

For BHs, an accurate approximation can be obtained in the \textit{pole-dipole approximation}, where only the first two terms in the gravitational skeleton are retained. This approximation is accurate by virtue of the no-hair theorems, which state that higher multipoles of BHs in GR are uniquely determined by only the mass monopole and spin dipole of the BH~\cite{Israel:1967wq, Israel:1967za, Carter:1971zc}. More recent studies in the context of EFT have focused on action principles~\cite{HANSON1974498, Bailey:1975fe, Porto:2005ac, Steinhoff:2014kwa} and symmetry generators~\cite{Steinhoff:2010zz}.

Within the post-Newtonian (PN) formalism~\cite{PW}, EFT has been shown to reproduce the equations of motion for spinning systems in a small spin expansion and to high PN order~\cite{Levi:2015ixa, Levi:2015uxa, Levi:2016ofk}, and has provided the means to re-sum these expressions at leading PN order~\cite{Siemonsen:2017yux}. Higher order multipoles, such as the spin-induced quadrupole and tidal deformations, have been calculated within the EFT formalism for NSs~\cite{Laarakkers:1997hb} and BHs~\cite{Thorne:1980rm}. This formalism has also been extended for NSs in scalar-tensor theories with scalarization~\cite{Sennett:2017lcx}. EFT is, thus, a powerful tool for modeling compact objects in relativistic field theories of gravity.

We here apply EFT methods to spinning BHs in dCS gravity. We work within an action formalism and derive a Lagrangian describing BHs as effective point particles in dCS gravity. To capture the additional degrees of freedom of BHs in dCS gravity, namely scalar dipole and spin-induced quadrupole moments, we work within the formalism of Eardley and Will~\cite{Will:1977zz} to promote the momenta of the BHs to functions of external fields, specifically the dCS scalar field and the Riemann tensor, respectively. We require our action to be invariant under a parity transformation and shift of the scalar field in addition to the usual symmetries of effective actions in GR. This allows us to explicitly fix the form of the Lagrangian up to two (constant) undetermined factors, namely $C_{\vartheta}^{\dcs}$ and $\delta C_{\QUAD}^{\dcs}$, for the scalar dipole and monopole-quadrupole interactions, respectively. These additional contributions to the matter action source both the dCS scalar field and the dCS metric perturbation. We match the near-zone (NZ) field solutions to those of isolated BHs in dCS gravity, fixing the undetermined factors. From this action, we derive the equations of motion and proceed with a PN expansion, ultimately obtaining the modified precession equations.

With the PN expansions of the precession equations at hand, we then solve them numerically and show that certain quantities remain conserved upon evolution. In particular, the total angular momentum of the system is conserved up to radiation-reaction effects. Such a result is important, as it allows us to choose a preferred coordinate system co-aligned with the direction of the total angular momentum, which results in simplifications of the dynamics~\cite{Chatziioannou:2016ezg, Chatziioannou:2017tdw}. Moreover, we prove that certain weighted combination of the spin magnitudes, with certain dCS corrections, is also conserved up to radiation-reaction effects. Such a conserved quantity provides an additional constant of the motion that is crucial when solving the orbital motion in quadrature~\cite{Racine:2008qv, Kesden:2014sla}. Finally, we solve the precession equations numerically and investigate their effect on the modulations of the GW amplitude. We find that the dCS corrections introduce modifications to these modulations that are proportional to the dCS coupling parameter. To our knowledge, this is the first calculation of spin-precession effects in modified gravity theories, and their inclusion in waveform modeling will be crucial to to test GR in the future. 

This paper, and its main results, are organized as follows. Section~\ref{prelim} presents a brief review of dCS gravity from an action formalism, and motivates the need for an EFT approach. Section~\ref{eft-sym} details the construction of an effective matter Lagrangian from symmetry principles, with the matter Lagrangian for BHs in dCS gravity given in Eq.~\eqref{eq:L-dCS}. Section~\ref{eft-var} provides the details necessary to vary the effective matter Lagrangian, with the matter stress energy tensor, scalar field effective source, and equations of motion, given in Eq.~\eqref{eq:Tmat-eff},~\eqref{eq:eff-source},~\eqref{eq:dpdt-dCS}, and~\eqref{eq:dSdt-dCS}, respectively. We provide NZ field solutions for the dCS scalar field and metric perturbation in Sec.~\ref{nz} and match these to the isolated BH case. We perform the PN expansion of the equations of motion in Sec.~\ref{pn}, with the spin evolution given by Eqs.~\eqref{eq:Sdot-GR}-\eqref{eq:Sdot-MQ}, and the orbital angular momentum evolution by Eqs.~\eqref{eq:Ldot-DD}-\eqref{eq:Ldot-MQ}. We prove the existence of conserved quantities necessary to construct waveforms in dCS gravity in Sec.~\ref{na}. Section~\ref{conclusion} concludes and points to future research. 

Throughout this paper, we use the following conventions: $G = 1 = c$, $(\mu, \nu, \rho, ...)$ are spacetime indices, $(i, j, k, ...)$ are purely spatial indices, $(A, B, C, ...)$ are spacetime indices in a body-fixed orthonormal frame, $(I, J, K, ...)$ are purely spatial indices in a body-fixed orthonormal frame, $(\textgoth{a}, \textgoth{b}, \textgoth{c}, ...)$ are internal indices used to encode non-GR fields, (...) stands for the symmetrization of indices, while $\left[...\right]$ stands for anti-symmetrization.

%%%%%%%%%%%%%%%%%%%%%%%%%%%%%%%%%%%
\section{Dynamical Chern-Simons Gravity and Compact Binaries}
\label{prelim}

We begin by considering the action in dCS gravity, which generally can be written as $S = S_{\gr} + S_{\CS} + S_{\vartheta} + S_{\MAT}$, where
\begin{align}
\label{eq:GR-action}
S_{\gr} &= \kappa \int d^{4}x \sqrt{-g} R \,,
\\
\label{eq:CS-action}
S_{\CS} &=  \alpha_{4} \int d^{4}x \sqrt{-g} \; \vartheta \; {^{\star}RR} \,,
\\
\label{eq:scalar-action}
S_{\vartheta} &=- \frac{\beta}{2} \int d^{4}x \sqrt{-g} \; \nabla_{\mu} \vartheta \nabla^{\mu} \vartheta\,,
\\
\label{eq:mat-action-x}
S_{\MAT} &= \int d^{4} x \sqrt{-g} {\cal{L}}_{\MAT}\,.
\end{align}
Here, $(g, R)$ are the determinant and Ricci scalar associated with the spacetime metric $g_{\mu \nu}$. The CS part of the action contains the Pontryagin density ${^{\star}RR}$, which is constructed from the Riemann tensor $R_{\mu \nu \rho \sigma}$ and its dual ${{^{\star}R}^{\mu}}_{\nu \rho \sigma} = (1/2) {\epsilon_{\rho \sigma}}^{\alpha \beta} {R^{\mu}}_{\nu \alpha \beta}$, with $\epsilon^{\mu \nu \rho \sigma}$ the Levi-Civita tensor. The field $\vartheta$ appearing in the CS action is a pseudo-scalar field, while $(\alpha_{4}, \beta)$ are coupling constants of the theory, and $\kappa = (16 \pi)^{-1}$~\cite{jackiw, Alexander:2009tp, quadratic}. Finally, ${\cal{L}}_{\MAT}$ is the matter Lagrangian density.

Variation of the action with respect to all fields produces the field equations, specifically
\begin{align}
\label{eq:dCS-field}
G_{\mu \nu} + \frac{\alpha_{4}}{2 \kappa} K_{\mu \nu} &= \frac{1}{2 \kappa} \left(T_{\mu \nu}^{\MAT} + T_{\mu \nu}^{\vartheta}\right)\,,
\\
\label{eq:box-theta}
\Box_{g} \vartheta &= - \frac{\alpha_{4}}{\beta} \: {^{\star}RR}\,,
\\
\label{eq:mat-cons}
\nabla_{\nu} T^{\mu \nu}_{\MAT} &= 0\,,
\end{align}
where $G_{\mu \nu}$ is the Einstein tensor, $T_{\mu \nu}^{\MAT}$ is the matter stress energy tensor, and
\begin{align}
K_{\mu \nu} &= -4 \left(\nabla^{\alpha} \vartheta\right) \epsilon_{\alpha \beta \gamma (\mu} \nabla^{\gamma} {R_{\nu)}}^{\beta} 
\nn \\
&\;\;\;\; + 4 \left(\nabla_{\alpha \beta} \vartheta\right) \; {^{\star}R}_{\mu \;\;\; \nu}^{\;\;\; \alpha \;\;\; \beta} \,,
\\
T_{\mu \nu}^{\vartheta} &= \beta \left(\nabla_{\mu} \vartheta \nabla_{\nu} \vartheta - \frac{1}{2} g_{\mu \nu} \nabla_{\alpha} \vartheta \nabla^{\alpha} \vartheta\right)\,.
\end{align}
In GR, ${\cal{L}}_{\MAT}$ only couples to the metric tensor ${\cal{L}}_{\MAT} = {\cal{L}}_{\MAT}(g_{\mu \nu})$. In dCS, however, strongly self-gravitating bodies are also sourced by the scalar field, which is a manifestation of the theory's violation of the strong equivalence principle; in the EFT approach, we find that one must consider ${\cal{L}}_{\MAT} = {\cal{L}}_{\MAT}(g_{\mu \nu}, \vartheta, R_{\mu \nu \rho \sigma})$, as we will see in the next section.

An additional generalization that could be considered is the presence of a potential in the scalar field's action $S_{\vartheta}$, specifically $V(\vartheta)$. The only restriction on the specific form $V(\vartheta)$ can take comes from the shift symmetry of the scalar field $\vartheta \rightarrow \vartheta + \textit{const.}$, which protects $\vartheta$ from acquiring a non-vanishing potential. Without a specific form for such a potential, obtaining solutions to the scalar field equation becomes intractable. Further, since dCS gravity arises as the low energy limit of certain high energy theories, the particular choice of the potential would need to be motivated from symmetries and/or physical scenarios of such theories~\cite{quadratic}. As such, we only consider the case when $V(\vartheta) = 0.$ The shift symmetry then arises directly from the fact that the Pontryagin density is a topologically invariant, specifically $\int d^{4}x \sqrt{-g} \; {^{\star}\!RR} = 0$, which leaves Eq.~\eqref{eq:CS-action} invariant under the shift symmetry.

The coupling constants in dCS gravity, $(\alpha_{4}, \beta)$, are dimensional, and thus it is useful to define a new coupling constant $\zeta= \xi/{\cal{M}}^{4}$, with $\xi = \alpha_{4}^{2}/\kappa \beta$ and ${\cal{M}}$ some representative length scale of the system. For isolated BHs, ${\cal{M}}$ corresponds to the BH's mass. Generally, we must consider solutions to the field equations in the limit $\zeta \ll 1$. If dCS gravity is treated as an exact theory, then the theory is likely to not have a well-posed initial value problem due to the presence of higher derivatives in the field equations~\cite{Delsate:2014hba}. It is thus necessary to treat dCS gravity as an effective theory, and consider solutions to the field equations that are small deformations from GR, specifically
\begin{equation}
\label{eq:metric-def}
g_{\mu \nu} = g_{\mu \nu}^{\gr} + \textgoth{h}_{\mu \nu}\,,
\end{equation}
where $\textgoth{h}_{\mu \nu} \sim {\cal{O}}(\zeta)$ is the dCS metric perturbation. An EFT treatment is also consistent with the idea of this theory emerging as the low-energy limit of a more fundamental UV completion of gravity. 

We are here interested in the inspiral of BH binaries due to GW emission.  We thus study the equations of motion within the PN framework, where the BHs are assumed to be slowly moving and the mutual gravitational interaction is weak, i.e. $v^{2} \ll 1 \gg m/r_{12}$, where $v$ is the relative orbital velocity, $m$ is the total mass and $r_{12}$ is the relative separation. Due to the virial theorem, $v^{2} \sim m/r_{12}$, and thus, we treat terms that are $v^{2n}$ as the same order as $(m/r_{12})^{n}$. The GR sector of the metric may be expanded as
\begin{equation}
g_{\mu \nu}^{\gr} = \eta_{\mu \nu} + h_{\mu \nu}\,,
\end{equation}
where to lowest order
\begin{align}
\label{eq:h00}
h_{00} &= 2 U + {\cal{O}}\left(c^{-4}\right)\,,
\\
\label{eq:h0j}
h_{0j} &= -4 U_{j} + {\cal{O}}\left(c^{-5}\right)\,,
\\
\label{eq:hjk}
h_{jk} &= 2 U \delta_{jk} + {\cal{O}}\left(c^{-4}\right)\,
\end{align}
for some potentials $U$ and $U_{i}$. A similar decomposition may be made for the dCS metric perturbation $\textgoth{h}_{\mu \nu}$.

We may now expand the field equations in both PN and dCS small deformations. Much of this has been done in~\cite{quadratic}, so we do not present all of the details here. The GR potentials $(U, U_{j})$ satisfy the field equations
\begin{align}
\label{eq:Box-U}
\Box_{\eta} U &= - 4 \pi \left(T^{00}_{\MAT} + T^{kk}_{\MAT}\right)\,,
\\
\label{eq:Box-Uj}
\Box_{\eta} U^{j} &= - 4 \pi T^{0j}_{\MAT}\,,
\end{align}
where $\Box_{\eta}$ is the flat space d'Alembertian operator. The dCS metric perturbation satisfies
\begin{equation}
\label{eq:h-dCS-eq}
\frac{\kappa}{2} \Box_{\eta} \textgoth{h}_{\mu \nu} = \alpha_{4} K_{\mu \nu} - \frac{1}{2} \delta T_{\mu \nu}^{\MAT} - \frac{1}{2} T_{\mu \nu}^{(\vartheta)}\,,
\end{equation}
where $\delta T_{\mu \nu}^{\MAT}$ is the perturbation of the stress energy tensor for matter. The evolution equations for the scalar field and the tensor $K_{\mu \nu}$ are given in terms of the GR metric perturbation $h_{\mu \nu}$ as
\begin{align}
\label{eq:scalar-evol-PN}
\Box_{\eta} \vartheta &= - \frac{2 \alpha_{4}}{\beta} \epsilon^{\alpha \beta \mu \nu} h_{\alpha \delta, \gamma \beta} {{h_{\nu}}^{[\gamma, \delta]}}_{\mu}\,,
\\
\label{eq:K-PN}
K_{\mu \nu} &= \vartheta^{,\delta}_{,\sigma} \eta_{\nu \alpha} \epsilon^{\alpha \sigma \beta \gamma} \left(h_{\mu [\gamma, \beta] \delta} + h_{\delta [\beta,\gamma] \mu}\right) 
\nn \\
&\;\;\;\;-2 \vartheta^{,\delta} \epsilon_{\delta \sigma \xi \mu} {{{h^{\sigma}}_{[\alpha}}^{,\alpha \xi}}\;_{\nu]}\,.
\end{align}

For a binary system, the PN expanded field equations can be simplified by noting that to lowest PN order, the GR potentials obey superposition, i.e., they may be written as $U = U_{1} + U_{2}$, where $U_{1,2}$ are the contributions from the individual bodies. In terms of these potentials, the evolution equation for the scalar field in Eq.~\eqref{eq:scalar-evol-PN} becomes
\begin{equation}
\label{eq:theta-eq}
\Box_{\eta} \vartheta = -32 \frac{\alpha_{4}}{\beta} \epsilon_{ijk} U_{,im} U_{k,jm}\,.
\end{equation}
Due to the non-linear behavior of the source term, the scalar field can be separated into two contributions: self terms, which only contain contributions from each individual body, and interaction (or cross) terms, where the source term contains potentials due to both bodies. In~\cite{quadratic}, it was shown that the self terms vanish. However, isolated BHs in dCS gravity possess a scalar field associated with a dipole moment. In order to accurately model BH binary systems in dCS gravity in a weak field approximation, an additional source term for the scalar dipole moment of the BHs must be added in the matter stress energy tensor, creating an additional source term in Eq.~\eqref{eq:theta-eq}. This was done in~\cite{quadratic}, but the addition of such a term also modifies the equations of motion for the individual BHs. Thus, it is necessary to develop a unified approach that accounts for the effective source term required for BH binaries and the modifications to the equations of motion in a more systematic way.

%%%%%%%%%%%%%%%%%%%%%%%%%%%%%%%%%%%
\section{EFT: Symmetries and Lagrangians}
\label{eft-sym}

Now that we have motivated the need to accurately account for extra degrees of freedom carried by BHs in dCS gravity, we present the details of using EFT to model BHs.

%-----------------------------------------------------------------------------
\subsection{EFT in GR}

We begin by reviewing the construction of an EFT in GR. In GR, astrophysical BHs obey the no hair theorems, and as a result, they are described by two quantities, the mass and the spin angular momentum. Thus, any EFT describing the motion of BHs in a background spacetime must include these two degrees of freedom. There are multiple methods to achieve such an EFT. Much of the early work focused on Mathisson's gravitational skeleton, a multipolar expansion of the stress energy tensor along a particle's wordline~\cite{Mathisson2010, Tulczyjew:1959}.   Alternatively, one can begin by constructing a worldline Lagrangian by considering underlying symmetries of the theory. We here take the latter approach and show that one obtains the same results as using the gravitational skeleton method. 

%-----------------------------------------------------------------------------
\subsubsection{Pole-Dipole Approximation}

First we define a few quantities associated with the particle. The particle moves along a worldline $z^{\mu}(\tau)$ with four velocity $u^{\mu} = dz^{\mu}/d\tau$. Here, $\tau$ is not the proper time of the particle, but some affine parameter of the worldline. We leave this as a general parameter in order to properly consider time reparameterization invariance of the action, and specialize it to the case of proper time at a later step. The particle moves in a background spacetime given by the metric $g_{\mu \nu}$. In order to account for the internal spin degrees of freedom, we define a body-fixed frame via an orthonormal tetrad $e_{A}^{\alpha}$, namely
\begin{equation}
\label{eq:e-def}
g_{\mu \nu} = \eta_{A B} e^{A}_{\mu} e^{B}_{\nu}\,, \qquad \eta_{A B} = g_{\mu \nu} e^{\mu}_{A} e^{\nu}_{B}\,.
\end{equation}
The rotation tensor associated with this body-fixed frame is then
\begin{equation}
\label{eq:Omega-def}
\Omega^{\mu \nu} = e^{\mu A} \frac{D e_{A}^{\nu}}{D\tau}
\end{equation}
where $D/D\tau = u^{\mu} \nabla_{\mu}$ is the covariant derivative with respect to $\tau$.

We are interested here in a Lagrangian for the worldlines of particles moving through spacetime. We may write the matter action as
\begin{align}
\label{eq:mat-action}
S_{\MAT}  = \int d\tau \L_{\MAT}(\tau)
\end{align}
where $\L_{\MAT}$ is the Lagrangian on the particle's worldline. This Lagrangian can be thought of as on-shell, and is related to the Lagrangian density via
\begin{equation}
{\cal{L}}_{\MAT}(x) = \int \frac{d \tau}{\sqrt{-g}} \L_{\MAT}(\tau) \delta^{(4)}\left[x - z(\tau)\right]\,.
\end{equation}

We now desire a specific form for $\L_{\MAT}(\tau)$. To obtain this, we consider a set of symmetries associated with theory, specifically 
\begin{itemize}
\item[(i)] \textit{reparameterization invariance}, 
\item[(ii)] \textit{Lorentz invariance}, and 
\item[(iii)] \textit{diffeomorphism invariance}. 
\end{itemize}
These symmetries are associated with the underlying physics being invariant regardless of the time, frame, and background that experiments are performed in, respectively. Further, these symmetries mandate that the Lagrangian must transform as a scalar under time reparameterizations, Lorentz transformations, and diffeomorphisms, respectively. 

We begin by considering the first of these. Under the transformation $\tau \rightarrow \lambda \tau$, with $\lambda$ a constant, the Lagrangian must transform as $\L_{\MAT} \rightarrow \lambda^{-1} \L_{\MAT}$ in order for the action to be invariant. The only variables associated with the worldline that transform in this way are $(u^{\mu}, \Omega^{\mu \nu})$. Thus, by applying Euler's theorem of homogeneous functions, the matter Lagrangian must be of the form
\begin{equation}
\label{eq:Lmat-eft}
\L_{\MAT}^{\podi} = p_{\mu} u^{\mu} + \frac{1}{2} S_{\mu \nu} \Omega^{\mu \nu}\,,
\end{equation}
where the momenta are defined by
\begin{equation}
p_{\mu} \equiv \frac{\partial \L_{\MAT}}{\partial u^{\mu}}\,, \qquad S_{\mu \nu} \equiv 2 \frac{\partial \L_{\MAT}}{\partial \Omega^{\mu \nu}}\,.
\end{equation}
The Lagrangian given in Eq.~\eqref{eq:Lmat-eft} is manifestly Lorentz invariant, and automatically restricted to the pole-dipole approximation. Applying the variational principle to this Lagrangian, which we detail in the next section, gives the equations of motion
\begin{align}
\frac{D p_{\mu}}{D\tau} &= \frac{1}{2} S_{\alpha \beta} {R^{\alpha \beta}}_{\gamma \mu} u^{\gamma}\,, \qquad \frac{D S_{\mu \nu}}{D\tau} &= 2 p_{[\mu} u_{\nu]}\,.
\label{eq:prec-eq-GR}
\end{align}

The spin tensor has a total of six non-zero components, but the spin vector only has three dynamical degrees of freedom. As a result, we must define a spin supplementary condition (SSC) to remove the non-dynamical degrees of freedom from the problem. A common choice is the covariant SSC, $S^{\mu \nu} p_{\nu} = 0$, which has the benefit of uniquely determining the particle's worldline given a set of initial data~\cite{Beiglbock:1965fun, Schattner:1979vp, Schattner:1979vn}. More general SSCs can be specified by $S^{\mu \nu} f_{\nu} = 0$, where $f^{\mu}$ is a suitably chosen time-like four-vector. Differentiating this condition with respect to $\tau$, using the precession equation in Eq.~\eqref{eq:prec-eq-GR}, and solving for the four-momentum results in the general relationship between $p_{\mu}$ and $u^{\mu}$
\begin{equation}
\label{eq:p-mom-eq}
p^{\mu} = \frac{1}{-f_{\alpha} u^{\alpha}} \left[\left(-f_{\nu} p^{\nu}\right) u^{\mu} + S^{\mu \nu} \frac{D f_{\nu}}{D\tau}\right]\,.
\end{equation}
For the covariant SSC, and working to leading order in spin, $p^{\mu} = m \, u^{\mu} + {\cal{O}}(S^{2})$, where $m^{2} = - p_{\mu} p^{\mu}$. 

With the above SSCs, we may define a covariant spin four-vector through
\begin{equation}
S_{\mu \nu} = \frac{1}{m} \epsilon_{\mu \nu \rho \sigma} p^{\rho} S^{\sigma}\,.
\end{equation}
However, we have not fully exhausted the freedom in the definition of spin. The spin four-vector still has one non-dynamical component that must be fixed. This is typically done by choosing $S_{\mu} f^{\mu} = 0$, which completes the SSC. It is important to note that different SSC conditions correspond to different gauges, but observable quantities are independent of the specific choice~\cite{Costa:2011zn}. We choose to work with the covariant SSC for the remainder of this work, i.e.~$S^{\mu \nu} p_{\nu} = 0$ and $S_{\mu} p^{\mu} = 0$.

What about diffeomorphism invariance? Consider an infinitesimal coordinate transformation\footnote{The four vector $\xi^{\mu}$ should not be confused with the dimensional dCS coupling constant $\xi$, which is a scalar.} $x^{\mu} \rightarrow x^{\mu} + \xi^{\mu}(x^{\alpha})$. The Lagrangian, and therefore the action, are scalars and thus invariant under such a transformation. However, the Lagrangian depends on tensors which do transform. As a result, these two properties produce a constraint on the Lagrangian's dependence on the metric. To derive this, we must find how the tensors within the problem transform under this diffeomorphism. For simplicity, we only consider the pole-dipole approximation, meaning the Lagrangian only depends on the dynamical variables $(z^{\mu}, e^{\alpha}_{A})$ through $(u^{\mu}, \Omega^{\mu \nu})$, and is assumed to be independent of derivatives of the metric. A straightforward calculation produces
\begin{align}
\label{eq:var-g}
\delta_{\xi^{\alpha}} g_{\mu \nu} &= - 2 g_{\beta (\mu} \partial_{\nu)} \xi^{\beta}\,,
\\
\label{eq:var-u}
\delta_{\xi^{\alpha}} u^{\mu} &= u^{\beta} \partial_{\beta} \xi^{\mu}\,,
\\
\label{eq:var-Omega}
\delta_{\xi^{\alpha}} \Omega^{\mu \nu} &= -2 \Omega^{\gamma [\mu} \partial_{\gamma} \xi^{\nu]}\,,
\end{align} 
where $\delta_{\xi^{\alpha}}$ stands for the variation operator due to an infinitesimal coordinate transformation. The variation of the matter Lagrangian now becomes
\begin{align}
\delta_{\xi^{\alpha}} \L_{\MAT} &= \frac{\partial \L_{\MAT}}{\partial u^{\mu}} \delta_{\xi^{\alpha}} u^{\mu} + \frac{\partial \L_{\MAT}}{\partial \Omega^{\mu \nu}} \delta_{\xi^{\alpha}} \Omega^{\mu \nu} + \frac{\partial \L_{\MAT}}{\partial g_{\mu \nu}} \delta_{\xi^{\alpha}} g_{\mu \nu}\,,
\\
&= p_{\mu} \delta_{\xi^{\alpha}} u^{\mu} + \frac{1}{2} S_{\mu \nu} \delta_{\xi^{\alpha}} \Omega^{\mu \nu} + \frac{\partial \L_{\MAT}}{\partial g_{\mu \nu}} \delta_{\xi^{\alpha}} g_{\mu \nu} \,.
\end{align}
Inserting the variations in Eqs.~\eqref{eq:var-g}-\eqref{eq:var-Omega}, we finally arrive at the constraint
\begin{equation}
p_{\mu} u^{\nu} + S_{\mu \alpha} \Omega^{\nu \alpha} - 2 \frac{\partial \L_{\MAT}}{\partial g_{\nu \alpha}} g_{\mu \alpha} = 0\,,
\end{equation}
which will become important when we derive the stress energy tensor for the particle. This completes our discussion of the pole-dipole approximation in GR.

%-----------------------------------------------------------------------------
\subsubsection{Quadrupole Approximation}

The preceding discussion, and the Lagrangian derived therein, is general in the sense that it applies both to BHs and NSs, but it is not the most general Lagrangian we may write down. In the case of NSs, spin-induced quadrupole and tidal deformations can play an important role in the dynamics of inspiraling binaries. For generic precessing binary systems, Ref.~\cite{Racine:2008qv} showed that the incorporation of the quadrupole-monopole interaction in the orbit-averaged precession equations results in a seventh constant of motion. This allows the precession equations to be solved in closed form~\cite{Kesden:2014sla}, which ultimately allowed for the creation of analytic Fourier-domain gravitational waveforms for generic precessing binary systems in GR~\cite{Chatziioannou:2016ezg, Chatziioannou:2017tdw}. Further, as we will show, quadrupole corrections play an important role in the dynamics of binary systems in dCS gravity. As such, we extend the pole-dipole approximation of the preceding section to include quadrupole corrections in the context of GR.

A more general Lagrangian that includes higher multipoles of our effective point particles may be obtained by allowing it to depend on additional external fields encoding spacetime curvature, beyond just the metric. A relatively simple way of incorporating these extra fields, while also preserving reparameterization and Lorentz invariance, is to promote the momenta to be functions of these extra fields. For example, if spin-induced quadrupole moments become important, then we can promote the four momentum to be,
\begin{align}
p_{\mu} \rightarrow {\cal{P}}_{\mu} &= p_{\mu} + \left(\frac{\partial {\cal{P}}_{\mu}}{\partial R_{\alpha \beta \gamma \delta}}\right)_{u,\Omega,g} \bigg|_{R_{\alpha \beta \gamma \delta} = 0} R_{\alpha \beta \gamma \delta} + ...\,,
\nn \\
&= p_{\mu} + n_{\mu} Q^{\alpha \beta \gamma \delta} R_{\alpha \beta \gamma \delta} + ...\,,
\end{align}
where ${\cal{P}}_{\mu}$ is now the canonical momentum, while $p_{\mu} := {\cal{P}}_{\mu}(R_{\alpha \beta \gamma \delta} = 0)$ will be referred to as the bare momentum. The quantity $n_{\mu}$ in the second line is a suitably chosen time-like vector, while the partial derivative is taken holding $(u^{\mu}, \Omega^{\mu \nu}, g_{\mu \nu})$ fixed. If we choose $n_{\mu} = a \; u_{\mu}$ with $a$ an undetermined coefficient, then
\begin{align}
\label{eq:cal-P-GR}
{\cal{P}}_{\mu} u^{\mu} = p_{\mu} u^{\mu} + a \left(u_{\mu} u^{\mu}\right) Q^{\alpha \beta \gamma \delta} R_{\alpha \beta \gamma \delta}
\end{align}
and we must choose $a$ such that the action is still reparameterization invariant\footnote{More generally, one could choose $n_{\mu} = a \; u_{\mu} + b_{\mu}$ where $b_{\mu} u^{\mu} = 0$. Alternatively, one could also choose $n_{\mu} = a' \; p_{\mu} + b'_{\mu}$ with $b'_{\mu} p^{\mu} = 0$. In a small spin expansion, these two choices are equivalent.}. Since $\L_{\MAT}^{\QUAD} \to \lambda^{-1} \L_{\MAT}^{\QUAD}$ when $\tau \to \lambda \tau$, we must then choose $a = (-u_{\mu} u^{\mu})^{-1/2}$ such that 
\begin{equation}
\label{eq:Lmat-GR}
\L_{\MAT}^{\QUAD} = p_{\mu} u^{\mu} + \frac{1}{2} S_{\mu \nu} \Omega^{\mu \nu} - \frac{1}{6} (-u_{\mu} u^{\mu})^{1/2} J^{\alpha \beta \gamma \delta} R_{\alpha \beta \gamma \delta}\,,
\end{equation}
where we have written $Q^{\alpha \beta \gamma \delta} = (1/6) J^{\alpha \beta \gamma \delta}$ to recover Dixon's quadrupole moment~\cite{Dixon} and the superscript Q is shorthand for the quadrupole approximation. 

The precise form of $J^{\alpha \beta \gamma \delta}$ (or $Q^{\alpha \beta \gamma \delta}$) depends on the scenario, but for spin induced mass quadrupoles,
\begin{equation}
\label{eq:J-GR}
J^{\alpha \beta \gamma \delta} = - \frac{3}{m^{3}}  C_{\QUAD} {\cal{P}}^{[\alpha} S^{\beta] \rho} {S^{[\gamma}}_{\rho} {\cal{P}}^{\delta]}\,.
\end{equation}
where $C_{\QUAD}$ is the quadrupole moment scalar. For BHs, $C_{\QUAD} = 1$~\cite{Thorne:1980rm}, while for NSs, its value depends on the equation of state~\cite{Laarakkers:1997hb}. In order to obtain the structure of the mass quadrupole in Eq.~\eqref{eq:J-GR}, we used the fact that the SSC specifies a local frame whose time direction is determined by ${\cal{P}}_{\mu}$. The quadrupole moment $J^{\alpha \beta \gamma \delta}$ can be decomposed into components that are orthogonal to ${\cal{P}}_{\mu}$, which specifically define the stress, flow, and mass quadrupole of the body~\cite{Ehlers1977, Steinhoff:2010zz}. Of these, only the mass quadrupole contributes to the gravitational field outside of compact objects. One can then perform a decomposition into an irreducible representation of SO(3) of each of these components in the local frame.
    
%-----------------------------------------------------------------
\subsection{Modified EFT for dCS Gravity}

We now consider how to extend the EFT of BHs to dCS gravity. In the latter, BHs have two modifications from GR. The first is scalar hair in the form of a scalar dipole moment, which sources the scalar field and is linearly related to the spin of the BHs, to lowest order in a small spin expansion. The second is a modification to the BH spacetime away from the Kerr metric. In a weak field expansion, such a modification resembles a quadrupole contribution to the Newtonian gravitational potential and is second order in the BH's spin. Any EFT we build must account for both of these modifications.

Let us consider the first of these, since we already considered how to account for quadrupole contributions in the previous section. Any modification that we create should preserve the underlying symmetries of the effective theory in GR, in this case, reparameterization invariance specifically. However, the scalar dipole must couple to the scalar field in the Lagrangian in order for it to act as a source term. One way of achieving such a modification is to promote the four momentum to become functions of the scalar field as well as the Riemann tensor, specifically 
\begin{align}
p_{\mu} &\rightarrow {\cal{P}}(u^{\mu}, \Omega^{\mu \nu}, g_{\mu \nu}, R_{\mu \nu \rho \sigma}, \Psi_{\textgoth{a}})\,,
%\\
%S_{\mu \nu} &\rightarrow {\cal{S}}_{\mu \nu}(u^{\mu}, \Omega^{\mu \nu}, g_{\mu \nu}, R_{\mu \nu \rho \sigma}, \Psi_{\textgoth{a}})\,,
\end{align}
where $\Psi_{\textgoth{a}} = (\vartheta, \nabla_{\mu} \vartheta, ..., \nabla_{N} \vartheta)$ with $N = \mu_{1} \; \mu_{2} \; ... \; \mu_{n}$ shorthand for multi-index notation. A similar method has already been used in the context of EFTs when modeling scalarized NSs in scalar tensor theories~\cite{Sennett:2017lcx}. We can Taylor expand this new canonical momentum about the scalar field and its first derivatives, specifically
\begin{align}
\label{eq:P-of-theta}
{\cal{P}}_{\mu}(\Psi_{\textgoth{a}}) &= p_{\mu} + \frac{1}{6} (-u_{\rho} u^{\rho})^{-1/2} u_{\mu} J^{\alpha \beta \gamma \delta} R_{\alpha \beta \gamma \delta} 
\nn \\
&\;\;\;\;+ \left(\frac{\partial {\cal{P}}_{\mu}}{\partial \vartheta}\right)_{u, \Omega, g, R, \nabla \vartheta} \bigg|_{\vartheta = 0} \vartheta 
\nn \\
&\;\;\;\;+ \left(\frac{\partial {\cal{P}}_{\mu}}{\partial \nabla_{\alpha} \vartheta}\right)_{u, \Omega, g, R, \vartheta} \bigg|_{\nabla_{\alpha} \vartheta = 0} \nabla_{\alpha} \vartheta\,,
%\label{eq:S-of-theta}
%{\cal{S}}_{\mu \nu}(\Psi_{\textgoth{a}}) &= S_{\mu \nu} + \left(\frac{\partial {\cal{S}}_{\mu \nu}}{\partial \vartheta}\right)_{u, \Omega, g, R, \nabla \vartheta} \bigg|_{\vartheta=0} \vartheta\
%\nn \\
%&\;\;\;\;+ \left(\frac{\partial {\cal{S}}_{\mu \nu}}{\partial \nabla_{\alpha} \vartheta}\right)_{u, \Omega, g, R, \vartheta} \bigg|_{\nabla_{\alpha} \vartheta=0} \nabla_{\alpha} \vartheta \,,
\end{align}
where again $p_{\mu} := {\cal{P}}_{\mu}(\vartheta = 0, \nabla_{\alpha} \vartheta = 0, R_{\alpha \beta \gamma \delta}=0)$ is the bare momentum. We truncate the expansion here since the first two multipoles of the scalar field would enter the equations of motion at lowest PN order relative to higher order moments. Further, isolated BHs in dCS gravity do not possess a scalar quadrupole moment, with the next lowest order moment beyond the dipole being the scalar octopole. This will enter the precession equations at higher PN order than what we consider here.
% and $S_{\mu \nu} := {\cal{S}}_{\mu \nu}(\vartheta = \nabla_{\alpha} \vartheta = R_{\alpha \beta \gamma \delta}=0)$ is the bare spin tensor.

We are now left with determining $(\partial {\cal{P}}_{\mu}/ \partial \vartheta)$ and $(\partial {\cal{P}}_{\mu}/ \partial \nabla_{\alpha} \vartheta)$.
%, $(\partial {\cal{S}}_{\mu \nu}/ \partial \vartheta)$, and $(\partial {\cal{S}}_{\mu \nu}/ \partial \nabla_{\alpha} \vartheta)$. 
To do this, we apply the same procedure we used in GR, namely requiring the action be invariant under particular transformations. In addition to requiring the action satisfy the usual GR symmetries, we require it also be invariant under a shift of the scalar field $\vartheta \rightarrow \vartheta + {\rm const.}$, and under a parity transformation. By working in the formalism of Eardley and Will~\cite{Will:1977zz}, we have already ensured that the action is reparameterization and Lorentz invariant. Shift invariance can be easily achieved by requiring $(\partial {\cal{P}}_{\mu} / \partial \vartheta) = 0$.
% = (\partial {\cal{S}}_{\mu \nu} / \partial \vartheta)$.

Parity invariance is more difficult to handle. In dCS, the scalar field is actually a pseudo-scalar field, and thus, odd under a parity transformation, specifically $\hat{P}[\vartheta] = - \vartheta.$ Typically, to understand parity transformations, one must foliate the spacetime with time-like hypersurfaces and perform purely spatial reflections on said hypersurfaces (see e.g.~\cite{Alexander:2017jmt}). Parity then becomes a slicing dependent transformation. We here use an alternative method, whereby we attach odd parity to two additional tensor quantities, namely the tetrad associated with the co-rotating frame and the Levi-Civita tensor,
\begin{equation}
\label{eq:parity-def}
\hat{P}[e_{I}^{\mu}] = - e_{I}^{\mu}\,, \qquad \hat{P}[\epsilon_{\mu \nu \rho \sigma}] = - \epsilon_{\mu \nu \rho \sigma}\,,
\end{equation}
where the index $I$ is purely spatial. This then implies that $({\cal{P}}_{\mu}, u^{\mu}, S_{\mu \nu}, \Omega_{\mu \nu})$ are parity even, and the parity of the GR action is preserved. We discuss this parity transformation in more detail in Appendix~\ref{parity}.

%Unfortunately, these symmetry considerations are not sufficient to completely specify Eq.~\eqref{eq:P-of-theta}, and thus, we require a few more considerations. First, the form of Eq.~\eqref{eq:P-of-theta} is akin to a multipole expansion, with the coefficients $(\partial {\cal{P}}_{\mu} / \partial \Psi_{\textgoth{a}})$ effective multipoles. Using that BHs in dCS gravity have a dipolar scalar field in Eq.~\eqref{eq:P-of-theta}, we may truncate the expansions after the terms proportional to $\nabla_{\alpha} \vartheta$. 
%%Further, the nature of the scalar field is akin to that of a dipolar electric field and the dipole moment associated with it should be of mass- (electric-) type, thus $(\partial {\cal{S}}_{\mu \nu} / \partial \nabla_{\alpha} \vartheta) = 0$.

We can now use these parity considerations to determine the form of $(\partial {\cal{P}}_{\mu} / \partial \nabla_{\alpha} \vartheta)$. First, notice that this dipole moment must be odd under parity, $\hat{P}[\nabla_{\alpha} \vartheta] = - \nabla_{\alpha} \vartheta$. Second, to leading order in spin, the scalar dipole moment of an isolated BH in dCS gravity is linearly proportional to the spin vector of the BH, with corrections to this entering at ${\cal{O}}(S^{3})$~\cite{alejo}. Thus, we are left with one choice that satisfies both of these constraints, specifically
\begin{equation}
\left(\frac{\partial {\cal{P}}_{\mu}}{\partial \nabla_{\alpha} \vartheta}\right)_{\Psi_{\textgoth{a}} = 0} = \frac{1}{2 m^{2}} C_{\vartheta}^{\dcs} {\epsilon_{\mu}}^{\alpha \nu \rho} S_{\nu \rho}\,,
\end{equation}
where $C_{\vartheta}^{\dcs}$ is an undetermined coefficient, which we refer to as the scalar dipole constant, and is dependent on the coupling constants of the theory. Thus, the canonical momentum is now
\begin{align}
\label{eq:can-mom}
{\cal{P}}_{\mu} &= p_{\mu} + \frac{1}{6} (-u_{\rho} u^{\rho})^{-1/2} u_{\mu} J^{\alpha \beta \gamma \delta} R_{\alpha \beta \gamma \delta} 
\nn \\
&+ \frac{1}{m^{2}} C_{\vartheta}^{\dcs} \; {{^{\star} S}_{\mu}}^{\alpha} \nabla_{\alpha} \vartheta\,,
\end{align}
where we have used ${^{\star}S}_{\mu \nu} = (1/2) {\epsilon_{\mu \nu}}^{\rho \sigma} S_{\rho \sigma}$. Beyond the scalar dipole constant, we have now fully exhausted the freedom in how our new Lagrangian depends on the dCS scalar field.

We still require additional contributions to our Lagrangian to account for the modified spacetime of BHs in dCS gravity. Fortunately, these quadrupole contributions take the same form as those in GR, the only difference being the overall undetermined factor, which now must depend on the coupling constants of dCS gravity. Thus, we finally arrive at the Lagrangian
\begin{align}
\label{eq:L-dCS}
\L_{\MAT}^{\dcs} &= p_{\mu} u^{\mu} + \frac{1}{2} S_{\mu \nu} \Omega^{\mu \nu} + \frac{1}{m^{2}} C_{\vartheta}^{\dcs} \; {{^{\star}S}_{\mu}}^{\alpha} u^{\mu} \nabla_{\alpha} \vartheta 
\nn \\
&- \frac{1}{6} (-u_{\mu} u^{\mu})^{1/2} J^{\alpha \beta \gamma \delta} R_{\alpha \beta \gamma \delta}\,,
\end{align}
where our ansatz for the mass quadrupole is now
\begin{equation}
\label{eq:J-dCS}
J^{\alpha \beta \gamma \delta} = -\frac{3}{m^{3}}\left(1 + \delta C_{\QUAD}^{\dcs}\right) {\cal{P}}^{[\alpha} S^{\beta] \rho} {S^{[\gamma}}_{\rho} {\cal{P}}^{\delta]}\,,
\end{equation}
with $\delta C_{\QUAD}^{\dcs}$ the dCS modification to the quadrupole moment scalar and we have restricted our attention to BHs by taking $C_{\QUAD}^{\gr} = 1$. 
%Notice that the canonical momentum ${\cal{P}}_{\mu}$ appears in our ansatz for the mass quadrupole, rather than the bare momentum $p_{\mu}$. This has to do with a different choice of SSC, namely
%
%\begin{equation}
%\label{eq:SSC-dCS}
%S^{\mu \nu} {\cal{P}}_{\nu} = 0 \,, \qquad S^{\mu} {\cal{P}}_{\mu} = 0\,.
%\end{equation}
%
%Finally, note the different signs in Eqs.~\eqref{eq:L-dCS}-\eqref{eq:J-dCS} as compared to Eqs.~\eqref{eq:Lmat-GR}-\eqref{eq:J-GR}. This is merely a convention that does not effect the dynamics of the system.

As a final point, we consider the diffeomorphism invariance of the Lagrangian given in Eq.~\eqref{eq:L-dCS}. The variations given in Eqs.~\eqref{eq:var-g}-\eqref{eq:var-Omega} still hold, but we must now consider the additional variations
\begin{align}
\delta_{\xi^{\alpha}} \left(\nabla_{\mu} \vartheta\right) &= - \partial_{\mu} \xi^{\beta} \nabla_{\beta} \vartheta\,,
\\
\delta_{\xi^{\alpha}} R_{\mu \nu \rho \sigma} &= 2 R_{\mu \nu \gamma [\rho} \partial_{\sigma]} \xi^{\gamma} + 2 R_{\rho \sigma \gamma [\mu} \partial_{\nu]} \xi^{\gamma}\,.
\end{align}
Following the same procedure for the effective Lagrangian in GR, we arrive at the constraint
\begin{align}
\label{eq:dLdg-dCS}
{\cal{P}}_{\mu} u^{\alpha} &- S_{\sigma \mu} \Omega^{\alpha \sigma} - 2 \frac{\partial \L_{\MAT}^{\dcs}}{\partial g_{\alpha \beta}} g_{\mu \beta} - \frac{1}{m^{2}} C_{\vartheta}^{\dcs} \; {{^{\star} S}_{\nu}}^{\alpha} u^{\nu} \nabla_{\mu} \vartheta 
\nn \\
&- \frac{2}{3} J^{\gamma \nu \sigma \alpha} R_{\gamma \nu \mu \sigma} = 0
\end{align}
This completes the derivations of the effective Lagrangian in dCS gravity.

%%%%%%%%%%%%%%%%%%%%%%%%%%%%%%%%%%%
\section{EFT: Variational Principle}
\label{eft-var}

We now consider the variation of the total action for dCS gravity, which is given by Eqs.~\eqref{eq:GR-action}-\eqref{eq:mat-action-x}. The variation of the Einstein-Hilbert, Chern-Simons, and scalar field actions, Eqs.~\eqref{eq:GR-action}-\eqref{eq:scalar-action}, have been previously explored in~\cite{quadratic} for example. We will not explain the variations of these terms explicitly here. On the other hand, the matter action now depends on additional fields beyond the metric, specifically the Riemann tensor and the dCS scalar field. We thus provide the details of the variational principle for the effective Lagrangian in Eq.~\eqref{eq:L-dCS}. 

We consider a manifestly covariant variation of the action with respect to $(u^{\mu}, \Omega^{\mu \nu}, g_{\mu \nu}, R_{\mu \nu \rho \sigma}, \nabla_{\mu} \vartheta)$ along the lines of~\cite{Steinhoff:2014kwa}. To do this, we must define a few things. First, we introduce the operator $\hat{\cal{G}}_{\mu}^{\nu}$, which is defined through the covariant and Lie derivative operators as
\begin{align}
\nabla_{\alpha} &= \partial_{\alpha} + \Gamma^{\mu}_{\nu \alpha} \hat{\cal{G}}^{\nu}_{\mu}\,,
\\
{\cal{L}}_{\xi^{\alpha}} &= \xi^{\mu} \partial_{\mu} - \left(\partial_{\nu} \xi^{\mu}\right) \hat{\cal{G}}^{\nu}_{\mu}\,.
\end{align}
This operator acts on spacetime indices, but not body-fixed indices. For example,
\begin{align}
\hat{\cal{G}}^{\nu}_{\mu} \phi &= 0\,,
\\
\hat{\cal{G}}^{\nu}_{\mu} V_{\beta} &= - \delta^{\nu}_{\beta} V_{\mu}\,,
\\
\hat{\cal{G}}^{\nu}_{\mu} T^{\alpha}_{\beta} &= - \delta^{\nu}_{\beta} T_{\mu}^{\alpha} + \delta^{\alpha}_{\mu} T_{\beta}^{\nu}\,,
\end{align}
where $\phi$ is a scalar, and $V_{\mu}$ and $T^{\nu}_{\mu}$ are first and second rank tensors, respectively. Using this linear operator, we can define the covariant differential and variation along the worldline $z^{\alpha}(\tau)$ by
\begin{align}
D &= d + \left(\Gamma^{\mu}_{\nu \alpha} dz^{\alpha}\right) \hat{\cal{G}}^{\nu}_{\mu}\,,
\\
\Delta &= \delta_{z} + \delta z^{\alpha} \nabla_{\alpha} = \delta + \Gamma^{\mu}_{\nu \alpha} \delta z^{\alpha} \hat{\cal{G}}^{\nu}_{\mu}\,,
\end{align}
where $\delta_{z}$ corresponds to intrinsic variations of a tensor field along the worldline and the term proportional to $\delta z^{\alpha}$ corresponds to shifts of the worldline with $\delta = \delta_{z} + \delta z^{\alpha} \partial_{\alpha}$. This is primarily relevant to the variation of the matter action, with the variations of the remaining parts of the total action following the usual prescription~\cite{CSreview}.
%Since the Lagrangian is a scalar, the covariant variation is equivalent to the scalar variation, i.e. $\Delta L_{\rm m} = \delta L_{\rm m}$, where $\delta = \delta_{z} + \delta z^{\alpha} \partial_{\alpha}$. However, the Lagrangian depends on tensors which must be varied in a covariant manner, just as was the case in the previous section when studying diffeomorphism invariance. 
Thus, the variation of the matter Lagrangian $\L_{\MAT}(u^{\mu}, \Omega^{\mu \nu}, g_{\mu \nu}, R_{\mu \nu \rho \sigma}, \nabla_{\mu} \vartheta)$ becomes
\begin{align}
\Delta \L_{\MAT}^{\dcs} &= {\cal{P}}_{\mu} \Delta u^{\mu} + \frac{1}{2} S_{\mu \nu} \Delta \Omega^{\mu \nu} + \frac{\partial \L_{\MAT}^{\dcs}}{\partial g_{\mu \nu}} \Delta g_{\mu \nu} 
\nn \\
&+ \frac{\partial \L_{\MAT}^{\dcs}}{\partial \nabla_{\mu} \vartheta} \Delta\left(\nabla_{\mu} \vartheta\right) + \frac{\partial \L_{\MAT}^{\dcs}}{\partial R_{\alpha \beta \gamma \delta}} \Delta R_{\alpha \beta \gamma \delta}\,.
\end{align}
We consider the variations of each of these terms individually.

First, consider the variations of the four velocity $\Delta u^{\mu}$. Expanding out the variation, we have
\begin{equation}
\Delta u^{\mu} = \delta \frac{d z^{\mu}}{d\tau} + \Gamma^{\mu}_{\alpha \beta} u^{\alpha} \delta z^{\beta} = \frac{D \delta z^{\mu}}{D \tau}\,,
\end{equation}
where we have used the fact that $u^{\alpha} \nabla_{\alpha} = D/D\tau$ and that the scalar variation $\delta$ commutes with ordinary derivatives with respect to $\tau$. Similarly, the variation of the scalar field follows the same procedure, giving,
\begin{equation}
\Delta \left(\nabla_{\mu} \vartheta\right) = \delta_{z} \left(\nabla_{\mu} \vartheta\right) + \delta z^{\alpha} \nabla_{\alpha \mu} \vartheta\,,
\end{equation}
where $\nabla_{\alpha \mu} = \nabla_{\alpha} \nabla_{\mu}$. Further, $\Delta g_{\mu \nu} = \delta_{z} g_{\mu \nu}$ by metric compatibility.

The variation of the rotation tensor, $\Delta \Omega^{\mu \nu}$, requires a significantly more in depth calculation. Consider first the commutators $[\hat{{\cal{G}}}^{\nu}_{\mu}, \hat{{\cal{G}}}^{\beta}_{\alpha}]$ and $[\Delta, D]$, which, after a lengthy but straightforward calculation, give
\begin{align}
[\hat{{\cal{G}}}^{\nu}_{\mu}, \hat{{\cal{G}}}^{\beta}_{\alpha}] &= \delta^{\beta}_{\mu} \hat{{\cal{G}}}^{\nu}_{\alpha} - \delta^{\nu}_{\alpha} \hat{{\cal{G}}}_{\mu}^{\beta}\,,
\\
[\Delta, D] &= \left(\delta_{z} \Gamma^{\mu}_{\nu \alpha} - \delta z^{\beta} {R^{\mu}}_{\nu \alpha \beta}\right) dz^{\alpha} \hat{{\cal{G}}}_{\mu}^{\nu}\,.
\end{align}
Now, recall that the tetrad specifying the co-rotating frame $e_{A}^{\mu}$ is not independent of the metric, but is related to it by Eq.~\eqref{eq:e-def}. Consider, $e_{A \mu} \Delta e^{\nu}_{A}$, which we split into symmetric and anti-symmetric parts to obtain
\begin{align}
e^{A \mu} \Delta e^{\nu}_{A} &= e^{A [\mu} \Delta e^{\nu]}_{A} + \frac{1}{2} \Delta \left(e^{A \mu} e^{\nu}_{A}\right)
\nn \\
&= \Delta \Theta^{\mu \nu} - \frac{1}{2} g^{\mu \alpha} g^{\nu \beta} \delta_{z} g_{\alpha \beta}
\end{align}
where we have defined $\Delta \Theta^{\mu \nu} \equiv e^{A [\mu} \Delta e^{\nu]}_{A}$. This term is anti-symmetric and thus independent from variations of the metric, which is the symmetric part of the above expression. We, thus, now have a way of extracting variations of the worldline degrees of freedom associated with $e_{A}^{\mu}$ independent of the metric. Applying these results, we find
\begin{align}
\Delta \Omega^{\mu \nu} &= \frac{D \Delta \Theta^{\mu \nu}}{D \tau} + 2 {\Omega_{\sigma}}^{[\mu} \Delta \Theta^{\nu] \sigma} + {R^{\mu \nu}}_{\alpha \beta} u^{\alpha} \delta z^{\beta} 
\nn \\
&+ \Omega^{\alpha [\mu} g^{\nu] \beta} \delta_{z} g_{\alpha \beta} + g^{\beta [\mu} g^{\nu] \rho} u^{\alpha} \nabla_{\beta} \delta_{z} g_{\rho \alpha}\,,
\end{align}
where we have made use of
\begin{equation}
\label{eq:Christoffel-var}
\delta_{z} \Gamma^{\mu}_{\alpha \beta} = \frac{1}{2} g^{\gamma \mu} \left(\nabla_{\alpha} \delta_{z} g_{\beta \gamma} + \nabla_{\beta} \delta_{z} g_{\alpha \gamma} - \nabla_{\gamma} \delta_{z} g_{\alpha \beta}\right)\,.
\end{equation}
This completes the derivation of $\Delta \Omega^{\mu \nu}$.

Finally, consider the variation of the Riemann tensor, which becomes
\begin{equation}
\Delta R_{\mu \nu \alpha \beta} = \delta_{z} R_{\mu \nu \alpha \beta} + \delta z^{\rho} \nabla_{\rho} R_{\mu \nu \alpha \beta}\,.
\end{equation}
The second of these two terms does not require any simplifications, while the first may be evaluated using
\begin{align}
\delta_{z} {R^{\sigma}}_{\nu \alpha \beta} &= \nabla_{\alpha} \delta_{z} \Gamma^{\sigma}_{\nu \beta} - \nabla_{\beta} \delta_{z} \Gamma^{\sigma}_{\nu \alpha}\,,
\nn \\
&= g^{\sigma \rho}\left(\nabla_{[\alpha | \nu} \delta_{z} g_{\rho | \beta]} + \nabla_{[\alpha \beta]} \delta_{z} g_{\rho \nu} 
\right.
\nn \\
&\left.
\;\;\;\;- \nabla_{[\alpha | \rho} \delta_{z} g_{\nu | \beta]}\right)\,,
\end{align}
where we have used Eq.~\eqref{eq:Christoffel-var} to obtain the second equality.

We now have all of the components we need to vary the matter action. What is left is to regroup terms into overall factors for $[\Delta z^{\mu}, \Delta \Theta^{\mu \nu}, \delta_{z} g_{\mu \nu}, \delta_{z}(\nabla_{\mu} \vartheta)]$. We do not explicitly provide all of the steps here, as the calculation is lengthy but straightforward. Terms that depend on total derivatives of these variations, specifically $D\delta z^{\mu}/D\tau$ and $D\Delta \Theta^{\mu \nu}/D\tau$, must be integrated by parts, producing terms dependent on $D{\cal{P}}_{\mu}/D\tau$ and $DS_{\mu \nu}/D\tau$, respectively, as well as boundary terms. These boundary terms may be set to zero by requiring that $\delta z^{\mu}(-\infty) = 0 = \delta z^{\mu}(+\infty)$ and $\Delta \Theta^{\mu \nu}(-\infty) = 0 = \Delta \Theta^{\mu \nu}(+\infty)$. Regrouping the remaining terms, we arrive at the total variation of the matter action, specifically
\begin{widetext}
\begin{align}
\label{eq:Lmat-var}
\delta S_{\MAT}^{\dcs} &= \int d^{4}x \int d\tau \Bigg\{ \left[{\cal{P}}^{\mu} u^{\nu} \delta^{(4)} - \nabla_{\alpha} \left(S^{\alpha \mu} u^{\nu} \delta^{(4)}\right) - \frac{2}{3} \nabla_{\alpha \beta}\left(J^{\mu \alpha \beta \nu} \delta^{(4)}\right) - \frac{1}{m^{2}} C_{\vartheta}^{\dcs} \; {{^{\star} S}_{\alpha}}^{\mu} \nabla^{\nu} \vartheta \; u^{\alpha} \delta^{(4)} 
\right.
\nn \\
&\left.
\;\;\;\;\;\;\;\;\;\;\;\;\;\;\;\;+ \frac{1}{3} {R_{\beta \alpha \gamma}}^{\mu} J^{\nu \gamma \alpha \beta} \delta^{(4)}  \right] \frac{\delta_{z} g_{\mu \nu}}{2} + \frac{1}{m^{2}} C_{\vartheta}^{\dcs} \; {{^{\star} S}_{\mu}}^{\nu} u^{\mu} \delta_{z} \left(\nabla_{\nu} \vartheta\right) \delta^{(4)}
\nn\\
&\;\;\;\;\;\;\;\;\;\;\;\;\;\;\;\;+ \left[{\cal{P}}_{[\mu} u_{\nu]} + \frac{1}{m^{2}} C_{\vartheta}^{\dcs} \; {{^{\star} S}_{\alpha [\mu}} \nabla_{\nu]} \vartheta \; u^{\alpha} + \frac{2}{3} R_{\alpha \beta \rho [\mu} {J_{\nu]}}^{\rho \beta \alpha} - \frac{1}{2} \frac{D S_{\mu \nu}}{D \tau}\right] \Delta \Theta^{\mu \nu} \delta^{(4)}
\nn \\
&\;\;\;\;\;\;\;\;\;\;\;\;\;\;\;\;+ \left[\frac{1}{2} S_{\alpha \beta} {R^{\alpha \beta}}_{\rho \mu} u^{\rho} + \frac{1}{m^{2}} C_{\vartheta}^{\dcs} \; {{^{\star} S}_{\alpha}}^{\beta} u^{\alpha} \nabla_{\mu \beta} \vartheta - \frac{1}{6} J^{\alpha \beta \gamma \delta} \nabla_{\mu} R_{\alpha \beta \gamma \delta} - \frac{D {\cal{P}}_{\mu}}{D \tau}\right] \delta z^{\mu} \delta^{(4)} \Bigg\}\,,
\end{align}
\end{widetext}
where $\delta^{(4)} = \delta^{(4)}[x^{\mu} - z^{\mu}(\tau)]$, and we have used Eq.~\eqref{eq:dLdg-dCS} to eliminate any terms containing $\partial L_{\MAT}^{\dcs}/\partial g_{\mu \nu}$, which also eliminates terms containing $\Omega^{\mu \nu}$ from the third line above (and in Eq.~\eqref{eq:dSdt-dCS}). This completes the variational principle for our effective action.

Before continuing, it is worth discussing what each of these terms mean physically. Naturally, the variation of the total action must vanish, not just the variation of the matter action separately. As a result, the first of these terms, which is proportional to $\delta_{z} g_{\mu \nu}$, becomes the matter stress energy tensor sourcing the gravitational field equations in Eq.~\eqref{eq:dCS-field}. Explicitly, we have
\begin{align}
\label{eq:Tmat-eff}
\sqrt{-g} T^{\mu \nu} &= \int_{-\infty}^{+\infty} d\tau \left[{\cal{P}}^{(\mu} u^{\nu)} \delta^{(4)} - \nabla_{\alpha} \left(S^{\alpha (\mu} u^{\nu)} \delta^{(4)}\right) 
\right.
\nn \\
&\left.
- \frac{2}{3} \nabla_{\alpha \beta}\left(J^{(\mu| \alpha \beta |\nu)} \delta^{(4)}\right) + \frac{1}{3} {R_{\beta \alpha \gamma}}^{(\mu} J^{\nu) \gamma \alpha \beta} \delta^{(4)}  
\right.
\nn \\
&\left.
- \frac{1}{m^{2}} C_{\vartheta}^{\dcs} \; {{^{\star} S}_{\alpha}}^{(\mu} \nabla^{\nu)} \vartheta \; u^{\alpha} \delta^{(4)} \right]\,,
\end{align}
which is identical to the results of~\cite{Steinhoff:2009tk}, but with extra terms to account for the dCS scalar field. The first three terms in the above stress energy tensor act as mass, spin, and quadrupole source terms respectively. The remaining two constitute interaction terms with the field variables $(g_{\mu \nu}, \nabla_{\mu} \vartheta)$.

The second term in Eq.~\eqref{eq:Lmat-var} depends on $\delta_{z} (\nabla_{\mu} \vartheta)$, thus entering the scalar field evolution equation in Eq.~\eqref{eq:box-theta}. This extra term acts as an effective dipolar source term for the scalar field, which may be written as a scalar density
\begin{equation}
\label{eq:eff-source}
\rho_{\vartheta} = - \frac{C_{\vartheta}^{\dcs}}{4 \pi \beta m^{2} \sqrt{-g}} \int d\tau \; {{^{\star} S}_{\mu}}^{\nu} u^{\mu} \; \nabla_{\nu} \delta^{(4)}\,,
\end{equation}
and the scalar field evolution equation is now
\begin{equation}
\label{eq:theta-eq-eff}
\Box_{g} \vartheta = - \frac{\alpha_{4}}{\beta} \; {^{\star}RR} - 4 \pi \rho_{\vartheta}\,.
\end{equation}
As we will show, in a PN expansion, this effective source term is identical to the one found in~\cite{quadratic}.

The remaining two terms in Eq.~\eqref{eq:Lmat-var}, which are proportional to $\delta z^{\mu}$ and $\Delta \Theta^{\mu \nu}$, must vanish independently, and constitute the equations of motion for the effective particle,
\begin{align}
\label{eq:dpdt-dCS}
\frac{D{\cal{P}}_{\mu}}{D\tau} &= \frac{1}{2} S_{\alpha \beta} {R^{\alpha \beta}}_{\rho \mu} u^{\rho} + \frac{1}{m^{2}} C_{\vartheta}^{\dcs} \; {{^{\star} S}_{\alpha}}^{\beta} u^{\alpha} \nabla_{\mu \beta} \vartheta 
\nn \\
&- \frac{1}{6} J^{\alpha \beta \gamma \delta} \nabla_{\mu} R_{\alpha \beta \gamma \delta}\,,
\\
\label{eq:dSdt-dCS}
\frac{DS_{\mu \nu}}{D\tau} &= 2 {\cal{P}}_{[\mu} u_{\nu]} + \frac{2}{m^{2}} C_{\vartheta}^{\dcs} \; {{^{\star} S}_{\alpha [\mu}} \nabla_{\nu]} \vartheta \; u^{\alpha} 
\nn \\
&+ \frac{4}{3} R_{\alpha \beta \rho [\mu} {J_{\nu]}}^{\rho \beta \alpha}\,.
\end{align}
In a binary system, the first of these determines how the acceleration of the particle is modified by the dCS perturbations, and ultimately determines the evolution equations for the orbital angular momentum. The second equation of motion is the spin-precession equation for the particle. This completes our discussion of an EFT for dCS gravity.

%%%%%%%%%%%%%%%%%%%%%%%%%%%%%%%%%%%
\section{Near-Zone Field Solutions}
\label{nz}

The equations of motion given in Eqs.~\eqref{eq:dpdt-dCS}-\eqref{eq:dSdt-dCS} are dependent on the fields $\vartheta$, $g_{\mu \nu}$, and $R_{\alpha \beta \gamma \delta}$. Thus, now that we have our effective matter action in hand, we must consider the fields generated by matter sources before we can expand the equations of motion. Since we are interested in the dynamics of binary systems, we restrict our attention to two bodies, and work to solve the field equations in a PN expansion within the NZ, where effects due to retarded time are higher PN order and we can take time derivatives to spatial derivatives via $\partial_{t} = -v^{k} \partial_{k}$. We work to second order in a small spin expansion. We desire to obtain the leading order terms in the spin-precession equations, which means we must work to leading PN order in all source terms. Finally, we take $u_{\mu} u^{\mu} = - 1$, thus fixing $\tau$ to be proper time.
%-----------------------------------------
\subsection{Scalar Field Evolution}

We begin by considering the evolution of the dCS scalar field in the NZ, which is governed by Eq.~\eqref{eq:theta-eq-eff}. The scalar field is sourced by two terms, one proportional to the Pontryagin density, and the other an effective source term. The contributions to the scalar field from the Pontryagin term in the NZ have been extensively studied in~\cite{quadratic}, which are higher PN order and only contain interaction terms. We thus do not consider these contributions here, and focus on the effective source term, which is given in Eq.~\eqref{eq:eff-source}. We present much of the machinery needed to evaluate NZ integrals in this section, as it is purely general and applies to any of the fields we  solve for.

By virtue of our choice of SSC, and the fact that we are working in a small spin expansion, we have that $S_{\mu \nu} u^{\mu} = {\cal{O}}(S^{3}) = S_{\mu} u^{\mu}$, and thus we may write
\begin{equation}
S_{\mu} = - \frac{1}{2} \epsilon_{\mu \nu \rho \sigma} u^{\nu} S^{\rho \sigma} = - {{^{\star} S}_{\mu \nu}} u^{\nu}\,.
\end{equation}
From the normalization condition for the four-velocity, we have
\begin{align}
u^{0} = 1 + (U + \delta U^{\dcs}) + \frac{1}{2} v^{2} + {\cal{O}}(c^{-4})\,,
\end{align}
where $\delta U^{\dcs}$ is the dCS correction to the Newtonian potential $U$, and $v^{2} = \delta_{ij} v^{i} v^{j}$ with $v^{i} = u^{i}/u^{0}$. Applying this to our SSC, we find $S_{0} = - S_{j} v^{j} + {\cal{O}}(c^{-3})$. Thus, terms proportional $S_{0}$ are in general higher PN order than terms proportional to $S_{j}$. Finally, from our linearization of the metric in Eq.~\eqref{eq:metric-def}, $(-g) = 1 - h - {\textgoth{h}}$. Applying all of this to Eq.~\eqref{eq:eff-source} and truncating at leading PN order, we have
\begin{equation}
\rho_{\vartheta} = - \frac{C_{\vartheta}^{\dcs}}{4 \pi \beta m^{2}} \int d\tau S^{k}(\tau) \partial_{k} \delta^{(4)}\left[x^{\mu} - z^{\mu}(\tau)\right]\,.
\end{equation}
As a final step of simplification, we can perform the integration over $\tau$ by exploiting the properties of the Dirac delta function to obtain
\begin{equation}
\label{eq:eff-source-PN}
\rho_{\vartheta} = - \frac{C_{\vartheta}^{\dcs}}{4 \pi \beta m^{2}} S^{k}(t) \partial_{k} \delta^{(3)}\left[\vec{x} - \vec{z}(t)\right]\,.
\end{equation}

The scalar field obeys the evolution equation $\Box_{\eta} \vartheta = -4 \pi \rho_{\vartheta}$, which has the general solution
\begin{equation}
\vartheta(t, \vec{x}) = \int d^{3}x' \frac{\rho_{\vartheta}\left(t - |\vec{x} - \vec{x}'|, \vec{x}'\right)}{|\vec{x} - \vec{x}'|} 
\end{equation}
Since we are working in the NZ, retardation effects are higher PN order, and we can simply write
\begin{align}
\vartheta(t, \vec{x}) &= \int d^{3}x' \frac{\rho_{\vartheta}\left(t, \vec{x}'\right)}{|\vec{x} - \vec{x}'|}\,,
\nn \\
&= - \frac{C_{\vartheta}^{\dcs}}{4 \pi \beta m^{2}} S^{k}(t) \int d^{3} x' \frac{\partial_{k} \delta^{(3)}\left[\vec{x}' - \vec{z}(t)\right]}{|\vec{x} - \vec{x}'|} \,,
\end{align}
where we have applied Eq.~\eqref{eq:eff-source-PN} to obtain the second equality. The Dirac delta function only depends on the difference $\vec{x} - \vec{z}(t)$, so we may replace the partial derivative acting on $\vec{x}'$ with a particle derivative $\tilde{\partial}_{k}$ that acts on $\vec{z}$, i.e., $\tilde{\partial}_{k} f(x^{\mu} - z^{\mu}) = - \partial_{k} f(x^{\mu} - z^{\mu})$ with $f(x^{\mu} - z^{\mu})$ an arbitrary functions of $x^{\mu} - z^{\mu}$. This particle derivative can then be pulled out of the integral, at which point the integration becomes trivial and we obtain
\begin{equation}
\vartheta(t,\vec{x}) = \frac{C_{\vartheta}^{\dcs}}{4 \pi \beta m^{2}} S^{k}(t) \tilde{\partial}_{k} \left(\frac{1}{|\vec{x} - \vec{z}(t)|}\right)\,.
\end{equation}
Finally, evaluating the derivative, we obtain
\begin{equation}
\label{eq:theta-sol-gen}
\vartheta(t,\vec{x}) = \frac{C_{\vartheta}^{\dcs}}{4 \pi \beta m^{2}} \frac{S^{k} n_{k}}{r^{2}}
\end{equation}
where $r = |\vec{x} - \vec{z}|$ and $\vec{n} = (\vec{x} - \vec{z})/r$.

At this stage, the solution in Eq.~\eqref{eq:theta-sol-gen} is purely general and must be matched to isolated BHs in dCS gravity to fix $C_{\vartheta}^{\dcs}$. Recall from~\cite{quadratic}, that the scalar field of an isolated BH is $\vartheta = (\vec{\mu} \cdot \vec{n})/r^{2}$, where $\vec{\mu} = (5 \alpha_{4}/2 \beta) \vec{\chi}$ is the scalar dipole moment of the BH with $\vec{\chi} = \vec{S}/m^{2}$ the reduced spin vector. Matching our result to this, we find
\begin{equation}
C_{\vartheta, {\BH}}^{\dcs} = 10 \pi \alpha_{4}\,,
\end{equation}
for BHs. The scalar field for the binary system is now just the sum of the individual contributions, specifically
\begin{equation}
\label{eq:theta-binary}
\vartheta = \frac{\vec{\mu}_{1} \cdot \vec{n}_{1}}{r_{1}^{2}} + \frac{\vec{\mu}_{2} \cdot \vec{n}_{2}}{r_{2}^{2}}\,,
\end{equation}
where $\vec{n}_{1,2} = \left(\vec{x} - \vec{z}_{1,2}\right)/r_{1,2}$. This completes the computation of the scalar field of the binary system in the NZ.

Before we continue, we would like to comment on the case of NSs in dCS gravity. The scalar dipole moment of NSs is dependent on the equation of state (EOS) of supra-nuclear matter, and thus, one would have to match the external scalar field to that of the interior to determine $C_{\vartheta, {\rm NS}}^{\dcs}$. A calculation along these lines was performed in~\cite{Yagi:2013mbt}, where fitting functions for the scalar dipole moment were computed for various EOSs. One could straightforwardly match said results to those here in order to determine $C_{\vartheta, {\rm NS}}^{\dcs}$. However, we are here only concerned with binary BHs, since these are expected to have non-negligible spins and thus present precession, so we do not consider this here.

%Neutron stars do not have a scalar dipole moment in dCS gravity, and the scalar field of a rotating NS is higher PN order than that shown in Eq.~\eqref{eq:theta-sol-gen}. Thus, for NSs, $C_{\vartheta, {\rm NS}}^{\dcs} = 0$. If the NS exists in a binary with a BH, then to leading PN order, the scalar field in Eq.~\eqref{eq:theta-binary} only has one contribution. On the other hand, if the NS exists in a binary with another NS, then Eq.~\eqref{eq:theta-binary} no longer applies, and one would likely need to go back to the effective matter Lagrangian to adequately account for the higher PN order effects of NSs.
%-----------------------------------------
\subsection{NZ Metric Perturbation}

We now consider the NZ solution for the metric perturbations. The GR metric perturbation $h_{\mu \nu}$ is governed by Eqs.~\eqref{eq:Box-U}-\eqref{eq:Box-Uj}, while the dCS metric perturbation satisfies Eq.~\eqref{eq:h-dCS-eq}. The GR metric perturbation is solely sourced by the GR sector of the matter stress energy tensor, and a detailed derivation of its explicit form can be found in~\cite{PW} for example. On the other hand, the dCS metric perturbation $\textgoth{h}_{\mu \nu}$ is sourced by the $K$-tensor in Eq.~\eqref{eq:K-PN}, the scalar field stress energy tensor $T_{\mu \nu}^{\vartheta}$, and the non-GR sector of the matter stress energy tensor $\delta T_{\mu \nu}^{\MAT}$. We can write the contributions from each of these terms as
\begin{align}
\textgoth{h}_{\mu \nu}^{K}(t, \vec{x}) &= 8 \alpha_{4} \int d^{3}x' \frac{K_{\mu \nu}(t, \vec{x}')}{|\vec{x} - \vec{x}'|}\,,
\\
\textgoth{h}_{\mu \nu}^{\vartheta}(t, \vec{x}) &= 4 \int d^{3}x' \frac{T_{\mu \nu}^{\vartheta}(t, \vec{x}')}{|\vec{x} - \vec{x}'|}\,,
\\
\label{eq:dCS-h-T}
\textgoth{h}_{\mu \nu}^{\MAT}(t, \vec{x}) &= 4 \int d^{3}x' \frac{\delta T_{\mu \nu}^{\MAT}(t, \vec{x}')}{|\vec{x} - \vec{x}'|}\,.
\end{align}
A simple PN counting of these terms reveals that the leading order contributions come from the matter stress energy tensor, so we focus our attention on it.

The full matter stress energy tensor is given in Eq.~\eqref{eq:Tmat-eff}. To begin, we must separate out the contributions to the matter stress energy tensor from GR and dCS gravity. 
%An explicit calculation of this is difficult to do, since the fields $g_{\mu \nu}$ and $R_{\mu \nu \rho \sigma}$ contain both GR and dCS pieces. Thus, we would need to expand out all repeated indices, separate out the pieces that are contracted only with $\eta_{\mu \nu}$, and those that are linear in both $h_{\mu \nu}$ and ${\textgoth{h}}_{\mu \nu}$. Fortunately, there are some simplifications we can immediately make, as they do not affect the end result for this work. First, we only desire to calculate the metric fields to leading PN order, so we may take repeated indices to be contracted with the flat space metric $\eta_{\mu \nu}$. Further, 
Since we are working in a linear approximation for the metric, we may write $R_{\mu \nu \rho \sigma} = {R}^{(0)}_{\mu \nu \rho \sigma}[h] + \delta R_{\mu \nu \rho \sigma}[\textgoth{h}]$, where $\delta R_{\mu \nu \rho \sigma}[\textgoth{h}]$ is the dCS perturbation. Second, the matter stress energy tensor contains the four-dimensional Dirac delta $\delta^{(4)}[x^{\mu} - z^{\mu}(\tau)]$. Since we are working with binary systems, the fields can be written as the sum of contributions from individual bodies, to the PN order we are working. As a result, contributions arising from the self interaction of fields with worldlines vanish via regularization. For example, the metric perturbation from the dipole-dipole interaction, which is the last line in Eq.~\eqref{eq:Tmat-eff}, becomes
\begin{align}
{\textgoth{h}}_{\mu \nu}^{\MAT, \dd} &= \frac{4}{m_{1}^{2}} C_{\vartheta}^{\dcs} \int_{-\infty}^{+\infty} d\tau \; {{{^{\star}S}_{1\;\mu}^{\alpha}}} u_{\alpha}^{1} \mu^{j}_{1} 
\nn \\
&\times \int d^{3}x' \delta'^{(4)}_{1} \partial'_{\nu j}\left(\frac{1}{|\vec{x}' - \vec{z}_{1}(\tau)|}\right) \frac{1}{|\vec{x} - \vec{x}'|}
\nn \\
&+ (1\leftrightarrow2) + \text{interaction terms} \,,
\end{align}
where $\delta'^{(4)}_{1} = \delta^{(4)}[x'^{\mu} - z_{1}^{\mu}(\tau)]$. If we replace the derivatives in the last integral with particle derivatives, we arrive at an integral of the form
\begin{equation}
\int d^{3}x' \frac{[\vec{x}' - \vec{z}_{1}(\tau)]^{N}}{|\vec{x}' - \vec{z}_{1}(\tau)|} \delta^{(3)}[\vec{x}' - \vec{z}_{1}(\tau)] = 0\,,
\end{equation}
which vanishes by regularization for all $N$. Thus, any contributions to Eq.~\eqref{eq:Tmat-eff} which contain fields mixed with particle momenta only contribute to the metric perturbation through interaction terms, which are higher PN order than what we are considering. We thus focus on the source terms, which are the first three terms in Eq.~\eqref{eq:Tmat-eff}.

Finally, we replace the canonical momentum ${\cal{P}}_{\mu}$ with the four velocity $u^{\mu}$. A straightforward calculation using the results of Appendix~\ref{p-to-u} shows that ${\cal{P}}_{\mu} = m \,u_{\mu}$ to the order we need here. Thus, we write the matter stress energy tensor as $\sqrt{-g} \, T^{\mu \nu} = \sqrt{-g} \, [{^{(0)}T}^{\mu \nu} + \delta T^{\mu \nu}]$, where
\begin{align}
\sqrt{-g} \; {^{(0)}T}^{\mu \nu} &= \int_{-\infty}^{+\infty} d\tau \left[m u^{\mu} u^{\nu} \delta^{(4)} - \nabla_{\alpha} \left(S^{\alpha (\mu} u^{\nu)} \delta^{(4)}\right)
\right.
\nn \\
&\left.
\hspace{50pt} \nabla_{\alpha \beta} \left(^{(0)}J^{(\mu| \alpha \beta|\nu)} \delta^{(4)}\right)\right]\,,
\\
\label{eq:delta-T}
\sqrt{-g} \, \delta T^{\mu \nu} &= -\frac{2}{3} \int_{-\infty}^{+\infty} d\tau \; \nabla_{\alpha \beta} \left(\delta J^{(\mu| \alpha \beta |\nu)} \delta^{(4)}\right)\,.
\end{align}
with $J^{\alpha \beta \gamma \delta} = \; ^{(0)}J^{\alpha \beta \gamma \delta} + \delta J^{\alpha \beta \gamma \delta}$, splitting the mass quadrupole into GR and dCS contributions. We may now solve the field equations given by Eqs.~\eqref{eq:Box-U}-\eqref{eq:Box-Uj} and~\eqref{eq:dCS-h-T}. The solutions for the GR potential have been extensively studied, for example in~\cite{PW}. We thus do not detail the derivation here, but the end result, to linear order in spin, is
\begin{align}
\label{eq:U-GR}
U(t, \vec{x}) &= \frac{m_{1}}{r_{1}} + \frac{3}{2} \frac{(\vec{n}_{1} \times \vec{v}_{1}) \cdot \vec{S}_{1}}{r_{1}^{2}}  - \frac{3}{2 m_{1}} \frac{S_{1}^{i} S_{1}^{j} n^{1}_{<ij>}}{r_{1}^{3}}  
\nn \\
&+ (1 \rightarrow 2)\,,
\\
\label{eq:Uj-GR}
U^{j}(t, \vec{x}) &= \frac{m_{1} v^{j}_{1}}{r_{1}} - \frac{1}{2} \frac{(\vec{n}_{1} \times \vec{S}_{1})^{j}}{r_{1}^{2}} + (1 \rightarrow 2)\,.
\end{align}

We now focus our attention on the dCS metric perturbation. Our ansatz for the spin-induced mass quadrupole moment is given in Eq.~\eqref{eq:J-dCS}. Applying the simplification procedure we used on the matter stress energy tensor, and inserting this into Eq.~\eqref{eq:delta-T}, we have
\begin{align}
\sqrt{-g} \delta T^{\mu \nu} &= \frac{2}{m} \delta C_{\QUAD}^{\dcs} u^{[\mu} S^{j] \rho} {S^{[k}}_{\rho} u^{\nu]} \partial_{jk} \delta^{(3)}[\vec{x} - \vec{z}(t)]\,.
\end{align}
By virtue of our choice of SSC and the small spin expansion, the components of the spin tensor are to leading PN order
\begin{equation}
S_{jk} = \varepsilon_{jki} S^{i}\,, \qquad S_{0j} = \varepsilon_{jik} v^{i} S^{k}\,,
\end{equation}
where $\varepsilon_{ijk}$ is the Levi-Civita symbol. As a result, the leading PN order contribution to the stress energy tensor perturbation is
\begin{align}
\delta T^{00} = - \frac{1}{2m} \delta C_{\QUAD}^{\dcs} \left[S^{j} S^{k} - (\vec{S} \cdot \vec{S}) \delta^{jk}\right] \partial_{jk} \delta^{(3)}[\vec{x} - \vec{z}(t)]\,,
\end{align}
where $\vec{S} \cdot \vec{S} = \delta_{mn} S^{m} S^{n}$. The procedure for solving for ${\textgoth{h}}_{\mu \nu}$ follows exactly the same steps as the previous section, and we eventually find the end result
\begin{equation}
\label{eq:U-dCS}
\delta U = - \frac{3}{2m} \delta C_{\QUAD}^{\dcs} S^{j} S^{k} \frac{n_{<jk>}}{r^{3}}\,,
\end{equation}
where we have written ${\textgoth{h}}_{00} = 2 \delta U$, and $n_{<jk>} = n_{j} n_{k} - (1/3) \delta_{jk}$ is the symmetric and trace-free projection of $n_{j} n_{k}$. Finally, we match this solution to the case of an isolated BH spacetime, described in harmonic coordinates in Appendix~\ref{harmonic}, and we find
\begin{equation}
\delta C_{\QUAD}^{\dcs} = - \frac{201}{112} \zeta\,,
\end{equation}
where $\zeta = \xi/ m^{4}$, with $\xi = \alpha_{4}^{2}/\kappa \beta$, is the dimensionless dCS coupling constant. Since we are linearizing in the metric perturbation, the potential of the binary is just the sum of individual contributions, specifically,
\begin{align}
\delta U &= \frac{201}{224} \zeta_{1} \chi_{1}^{jk} n^{1}_{<jk>} \left(\frac{m_{1}}{r_{1}}\right)^{3} + (1 \rightarrow 2)\,.
\end{align}
This completes the derivation of the NZ metric.

As a final point, note that there is also a gravitomagnetic component to the metric perturbation, specifically $\textgoth{h}_{0j} = -4 \delta U_{j}$, due to the mass quadrupole moment which we do not consider here. This contribution is not present in the isolated BH metric, since that metric is computed in a co-moving frame. Further, this term is suppressed by one factor of the velocity of the body, and thus it enters at higher PN order in the equations of motion as we will show in the next section. This is different from GR, where the spin contributions to both the Newtonian potential and gravitomagnetic potential enter at the same PN order in the precession equations, due to the scaling of the potentials with velocity. 
%{\ny{Wait, but you should also comment on the gravitomagnetic part o the isolated BH metric, which is linear in spin and independent of velocity. We don't include this term because it's much higher PN order, but you should say so here.}} 

This is not the only contribution to the gravitomagnetic sector for BHs. In the isolated case, the metric component $g_{t \phi}$ is linear in the BH's spin. However, this term is higher PN order than those that enter the $g_{00}$ or $g_{ij}$ components of the metric. This can be seen in Appendix~\ref{harmonic} when we transform the isolated BH metric from Boyer-Lindquist coordinates to harmonic coordinates. The gravitomagnetic sector is actually 2PN order higher than the leading order contribution from the potential $U$ given in Appendix~\ref{harmonic}. Since we are only working to leading PN order, we thus do not consider this here.
 
%%%%%%%%%%%%%%%%%%%%%%%%%%%%%%%%%%%
\section{Expansion of the Equations of Motion}
\label{pn}

Now that we have the solutions to the potentials in the NZ of the binary, we can focus on the PN expansion of the equations of motion given in Eq.~\eqref{eq:dpdt-dCS}-\eqref{eq:dSdt-dCS}.

%---------------------------------------------------------
\subsection{Evolution of the Spin Angular Momentum}

While we could focus on the precession equation given in Eq.~\eqref{eq:dSdt-dCS}, the spin tensor contains non-dynamical degrees of freedom. We thus begin by deriving the precession equations for the spin vector $S^{\mu}$, which from our covariant SSC above Eq.~\eqref{eq:p-mom-eq} is given by
\begin{align}
S^{\mu} &= - \frac{1}{2 {\cal{M}}} \epsilon^{\mu \nu \rho \sigma} {\cal{P}}_{\nu} S_{\rho \sigma}\,.
\end{align}
To derive the precession equation for $S^{\mu}$, we simply have to apply $D/D\tau = u^{\alpha} \nabla_{\alpha}$ to this expression. Doing so generates three terms, which are proportional to $D{\cal{M}}/D\tau$, $D{\cal{P}}_{\mu}/D\tau$, and $DS_{\rho \sigma}/D\tau$, respectively. Using Eq.~\eqref{eq:can-mom} and the fact that ${\cal{M}}^{2} = - {\cal{P}}_{\alpha} {\cal{P}}^{\alpha}$, a straightforward expansion shows that the first of these only generates terms that are ${\cal{O}}(S^{3})$, as we show in Appendix~\ref{p-to-u}. For the term proportional to $D{\cal{P}}_{\mu}/D\tau$, we use Eq.~\eqref{eq:dpdt-dCS}. The only term that contributes from this equation is the one proportional to the Riemann tensor, since the others introduce terms that are ${\cal{O}}(S^{3})$. The Riemann tensor in this term can be further expanded into GR and dCS pieces via $R_{\mu \nu \rho \sigma} = {^{(0)} R}_{\mu \nu \rho \sigma}[h] + \delta R_{\mu \nu \rho \sigma}[\textgoth{h}]$. The dCS term is already ${\cal{O}}(S^{2})$ as can be seen from Eq.~\eqref{eq:U-dCS}, and thus contributes to the precession equation at ${\cal{O}}(S^{3})$. The GR term enters the GR part of the precession equation at higher PN order, so we neglect it here. We are thus left with the contribution from $DS_{\rho \sigma}/D\tau$, which leads to
\begin{align}
\label{eq:dSvecdt-dCS}
\frac{DS^{\mu}}{D\tau} &= -\frac{1}{m^{3}} C_{\vartheta}^{\dcs} \epsilon^{\mu \alpha \rho \sigma} p_{\rho} S_{\sigma} \nabla_{\alpha} \vartheta 
\nn \\
&- \frac{2}{3 m} \epsilon^{\mu \nu \rho \sigma} p_{\nu} {^{(0)} R}_{\alpha \beta \gamma \rho}[h] {J_{\sigma}}^{\gamma \beta \alpha}\,,
\end{align}
where we have performed an expansion in small dCS coupling. One can show from Eq.~\eqref{eq:dSvecdt-dCS} that the magnitude of the spin four vector $S^{2} = S_{\mu} S^{\mu}$ is conserved, as we show in Appendix~\ref{p-to-u}.

As we have noted during our earlier discussion of SSCs, there are only three dynamical degrees of freedom in the spin four-vector $S^{\mu}$. The choice of SSC and frame causes these three dynamical degrees of freedom to be spread over all four components of $S^{\mu}$. It is thus useful to define a frame where the dynamical degrees of freedom are only the spatial components of the spin vector, i.e. $\bar{S}^{\mu} = (0, \bar{S}^{j})$. To do this, we follow~\cite{Damour:2007nc} and begin by considering the conserved norm of the spin four-vector, specifically
\begin{equation}
S^{2} = S_{\mu} S^{\mu} = G^{ij} S_{i} S_{j}
\end{equation}
where we have used our choice of SSC to write
\begin{equation}
G^{ij} = g^{ij} - 2 g^{0(i} v^{j)} + g^{00} v^{i} v^{j}\,.
\end{equation}
We now define the square root matrix of $G^{ij} = H^{ik} {H_{k}}^{j}$, which acts as a projection operator on the spin vector such that $\bar{S}_{j} = {H_{j}}^{k} S_{k}$. By doing this, we have projected the spin four-vector into a frame where the conserved spin norm is only related to the spatial components of the spin vector, i.e. $S^{2} = \delta^{ij} \bar{S}_{i} \bar{S}_{j}$. By working in a PN expansion, it is easy to show that
\begin{align}
\label{eq:H-proj}
H^{ij} &= \delta^{ij} \left(1 - \tilde{U} \right) - \frac{1}{2} v^{i} v^{j} + {\cal{O}}\left(\frac{1}{c^{4}}\right)
\end{align}
where $\tilde{U} = U + \delta U$.

The goal is now to derive the precession equation for $\bar{S}_{j}$. Returning to Eq.~\eqref{eq:dSvecdt-dCS}, we begin by focusing on the left-hand side of this equation. Expanding out the covariant time derivative, we find
\begin{align}
\label{eq:covd-to-normd}
\frac{DS_{j}}{D\tau} &= \frac{dS_{j}}{dt} - \Gamma^{\beta}_{\alpha j} u^{\alpha} S_{\beta}
\nn \\
&= \frac{dS_{j}}{dt} - {V_{j}}^{k} S_{k}
\end{align}
where
\begin{equation}
\label{eq:V-unbar}
{V_{j}}^{k} = \Gamma^{k}_{j0} + \Gamma^{k}_{ji} v^{i} - \Gamma^{0}_{0j} v^{k} - \Gamma^{0}_{ij} v^{i} v^{k}\,.
\end{equation}
Solving for $dS_{j}/dt$ using Eq.~\eqref{eq:dSvecdt-dCS}, we obtain
\begin{align}
\label{eq:S-dot-unbar}
\frac{dS_{j}}{dt} &= {V_{j}}^{k} S_{k} + \frac{1}{m^{2}} C_{\vartheta}^{\dcs} \varepsilon^{jik} S_{i} \partial_{k} \vartheta 
\nn \\
&+ \frac{1}{m} \left(1 + \delta C_{\QUAD}^{\dcs}\right) \varepsilon^{jik} S^{m} S_{k} \partial_{mi} U\,,
\end{align}
where we have PN expanded the right-hand side of Eq.~\eqref{eq:dSvecdt-dCS} using the fact that $\nabla_{\alpha} = \partial_{\alpha}$ and $^{(0)}R_{0j0k} = -\partial_{jk} U$ at leading PN order. We can now use this expression in the time derivative of the definition of $\bar{S}_{j}$ 
\begin{align}
\frac{d\bar{S}_{j}}{dt} &= {H_{j}}^{k} \dot{S}_{k} + {\dot{H}_{j}}^{k} S_{k}
\end{align}
to finally arrive at
\begin{align}
\frac{d\bar{S}_{j}}{dt} &= \bar{V}_{jk} \bar{S}^{k} + \frac{1}{m^{2}} C_{\vartheta}^{\dcs} {H_{j}}^{m} {\epsilon_{m}}^{ik} (H^{-1})_{in} \bar{S}^{n} \partial_{k}\vartheta
\nn \\
&+\frac{1}{m} \left(1 + \delta C_{\QUAD}^{\dcs}\right) {H_{j}}^{m} {\epsilon_{n}}^{ik} {(H^{-1})^{n}}_{p} 
\nn \\
&\times \bar{S}^{p} (H^{-1})_{kq} \bar{S}^{\QUAD} \partial_{mi}U
\end{align}
where we have defined
\begin{align}
{\bar{V}_{j}}^{k} = {H_{j}}^{m} {V_{m}}^{n} (H^{-1})_{n}^{\;\;k} + {\dot{H}_{j}}^{\;m} (H^{-1})_{m}^{\;\;\;k}\,,
\end{align}
and we have used that $S_{j} = {(H^{-1})_{j}}^{k} \bar{S}_{k}$, with $(H^{-1})^{ij}$ the inverse of $H^{ij}$. We recognize this as the PN expanded precession equations in dCS gravity. 

Before we continue, it is useful to evaluate $\bar{V}_{jk}$ in terms of the metric potentials $(U, U_{j}, \delta U)$. By applying Eqs.~\eqref{eq:H-proj} and~\eqref{eq:V-unbar}, we obtain to leading PN order
\begin{align}
\bar{V}_{jk}^{\gr} &= - 4 \partial_{[j} U_{k]} - 3 v_{[j} \partial_{k]} U\,,
\\
\bar{V}_{jk}^{\dcs} &= - 3 v_{[j} \partial_{k]} \delta U\,.
\end{align}
Note that the dCS perturbations to the Newtonian potential $\delta U$ are second order in spin, and thus $\bar{V}_{jk}^{\dcs}$ creates terms that are third order in spin in the precession equations. Therefore, to the order in spin we are working, we neglect these contributions. 

What is left now is to apply this to a BH binary system by inserting the potentials in Eqs.~\eqref{eq:U-GR}-\eqref{eq:Uj-GR}, and~\eqref{eq:theta-binary} into the above precession equation. Since we are dealing with a binary system, the potentials contain contributions from both bodies. Thus, we must regularize the contributions at the locations of each particle, or BH. For the first BH (labeled 1), the potentials then only contain contributions from the other body (labeled 2). Further, we drop the bars that appear on all of the spin vector for convenience. Applying the spatial derivatives, we find $\dot{\vec{S}}_{1} = \dot{\vec{S}}_{1, {\gr}} + \dot{\vec{S}}_{1, {\dd}} + \dot{\vec{S}}_{1, {\MQ}}$, where
\begin{align}
\label{eq:Sdot-GR}
\dot{\vec{S}}_{1, {\gr}} &= \frac{\mu}{r_{12}^{2}} \left(2 + \frac{3}{2} \frac{m_{2}}{m_{1}}\right) \left(\vec{n}_{12} \times \vec{v}_{12}\right) \times \vec{S}_{1}
\nn \\
&+\frac{1}{r_{12}^{3}} \left[3 \left(\vec{S}_{2} \cdot \vec{n}_{12}\right) \vec{n}_{12} - \vec{S}_{2}\right] \times \vec{S}_{1}
\nn \\
&+ \frac{3}{r_{12}^{3}} \frac{m_{2}}{m_{1}} \left(\vec{S}_{1} \cdot \vec{n}_{12}\right) \left(\vec{n}_{12} \times \vec{S}_{1}\right)\,,
\\
\label{eq:Sdot-DD}
\dot{\vec{S}}_{1, {\dd}} &= \left(\frac{25}{16}  \zeta_{12}\right) \frac{1}{r_{12}^{3}} \left[3 \left(\vec{S}_{2} \cdot \vec{n}_{12}\right) \vec{n}_{12} - \vec{S}_{2}\right] \times \vec{S}_{1}\,,
\\
\label{eq:Sdot-MQ}
\dot{\vec{S}}_{1, {\MQ}} &= \left(-\frac{201}{112}  \zeta_{1}\right) \frac{3}{r_{12}^{3}} \frac{m_{2}}{m_{1}}  \left(\vec{S}_{1} \cdot \vec{n}_{12}\right) \left(\vec{n}_{12} \times \vec{S}_{1}\right)\,,
\end{align}
with $\mu = m_{1} m_{2}/(m_{1} + m_{2})$, $\zeta_{1} = \xi/m_{1}^{4}$, $\zeta_{12} = \xi/m_{1}^{2} m_{2}^{2}$, $\vec{r}_{12} = \vec{z}_{2}(t) - \vec{z}_{1}(t)$, $\vec{n}_{12} = \vec{r}_{12}/r_{12}$, and $\vec{v}_{12} = d\vec{r}_{12}/dt$. The first term is the leading PN order spin-orbit, spin-spin, and quadrupole-monopole interactions in GR, respectively. The second term is the scalar dipole-dipole interaction, which takes the same form as the spin-spin contribution in GR, but with a different prefactor that is proportional to $\zeta_{12}$. The final term is the dCS monopole-quadrupole interaction, which takes the same form as the quadrupole-monopole interaction in GR, but with a different prefactor proportional to $\zeta_{1}$. Notice that these last two terms, which are unique to dCS gravity, are the same PN order as the leading spin-spin contribution to GR, and are thus leading PN order. To obtain the precession equations for the spin of the other BH, one simply needs to make the replacement $1 \leftrightarrow 2$ in the above equations. This completes the derivation of the spin-precession equations in dCS gravity.

%-----------------------------------------------------
\subsection{Evolution of the Orbital Angular Momentum}

We now turn our attention to the evolution equation for the orbital angular momentum. To obtain this, we must first find the relative acceleration between the two BHs, and extract the spin-dependent terms of the forces acting on the binary. Thus, we return to the evolution equation for the canonical momentum given in Eq.~\eqref{eq:dpdt-dCS}. Performing a PN expansion of this equation, and many of the consideration that go into this process, have already been detailed in the previous section. 

The only complications arise from converting the covariant time derivative to an ordinary one, as was done in Eq.~\eqref{eq:covd-to-normd}. In order to obtain the full expression for the angular momentum evolution in GR, one needs to work to 1PN order, which amounts to including the gravitomagnetic potential and nonlinear interactions of potential and velocities in the relative acceleration. This calculation is complicated, but can straightforwardly be done following the steps of the previous section, and is detailed in~\cite{PW}, for example. The end result of such a computation is
\begin{equation}
\frac{d v^{j}}{dt} = {\cal{F}}_{\Gamma, {\gr}}^{j} + {\cal{F}}_{\Gamma, {\dcs}}^{j} + {\cal{F}}_{\dd}^{j} + {\cal{F}}_{\MQ}^{j}
\end{equation}
where
\begin{align}
\label{eq:f-GR}
{\cal{F}}_{\Gamma, {\gr}}^{j} &= \partial^{j} U + \left(v^{2} - 4 U\right) \partial^{j} U - v^{j} v^{k} \partial_{k} U 
\nn \\
&- 4 v_{k} \partial^{j} U^{k}
\\
\label{eq:f-dCS}
{\cal{F}}_{\Gamma, {\dcs}}^{j} &= \partial^{j} \delta U\,,
\\
\label{eq:f-DD}
{\cal{F}}_{\dd}^{j} &= \frac{1}{m^{3}} C_{\vartheta}^{\dcs} \bar{S}_{i} \partial^{ji} \vartheta\,,
\\
\label{eq:f-MQ}
{\cal{F}}_{\MQ}^{j} &= - \frac{1}{2 m} \left(1 + \delta C_{\QUAD}^{\dcs}\right) \left(\bar{S}_{i} \bar{S}_{k} \partial^{jik} U - s^{2} \partial^{j} \nabla^{2} U\right)\,.
\end{align}
The GR force ${\cal{F}}^{j}_{\Gamma, {\gr}}$ needs to be split between contributions that do not contain the spins of the bodies, and those that do. From the former, the first term in Eq.~\eqref{eq:f-GR} is the Newtonian gravitational force, while the remaining terms are all 1PN order, and a conserved (neglecting precession) orbital angular momentum can be derived from these. The latter induce precession of this angular momentum. The spin contributions from all of the terms in Eq.~\eqref{eq:f-GR} coming from both the $U$ and $U^{k}$ potentials are all the same PN order.

Interestingly, this same peculiarity in the PN counting of spin terms in GR does not occur for dCS gravity, at least at the PN order we are working. The gravitomagnetic potential $\delta U^{j}$ associated with the dCS quadrupole moment is suppressed relative to $\delta U$ by a factor of ${\cal{O}}(v^{2})$. Thus, when it enters the acceleration given above, it is truly a 1PN higher order correction to the leading order contribution given in Eq.~\eqref{eq:f-dCS} and we neglect it here.

The above forces must be suitably regularized at the location of each BH, just as we did in the spin-precession equations. The relative acceleration can then be computed using $\vec{a} = d\vec{v}_{2}/dt - d\vec{v}_{1}/dt = d\vec{v}_{12}/dt$, which gives
\begin{align}
\label{eq:rel-a}
a^{j}&= \Delta {\cal{F}}_{\Gamma, {\gr}}^{j} + \Delta {\cal{F}}_{\Gamma, {\dcs}}^{j} + \Delta {\cal{F}}_{\dd}^{j} + \Delta {\cal{F}}_{\MQ}^{j}
\end{align}
where $\Delta {\cal{F}} = {\cal{F}}_{2} - {\cal{F}}_{1}$. When considering the GR force $\Delta {\cal{F}}_{\Gamma, {\gr}}^{j}$, and using the method of undetermined coefficients~\cite{PW} for example, the conserved angular momentum at 1PN order is given by
\begin{equation}
\label{eq:L-def}
\vec{L} = \mu \left(\vec{r}_{12} \times \vec{v}_{12}\right) \left[1 + \frac{1}{2} (1 - 3 \eta) v_{12}^{2} + (3 + \eta) \frac{M}{r_{12}}\right]
\end{equation}
where $M = m_{1} + m_{2}$ and recall that this is only conserved in the absence of precession. To obtain the angular momentum evolution, one simply has to take the time derivative of the above expression, and replace any instance of the relative acceleration with Eq.~\eqref{eq:rel-a}. The non-spinning contributions cancel, since the above angular momentum is conserved at that level. For simplicity, we only detail the calculation for the non-GR terms from this point forward. The derivation of the GR contributions are rather lengthy, but can be found in~\cite{PW, Bohe:2012mr}. 

For the non-GR terms, it suffices to consider only the Newtonian part of Eq.~\eqref{eq:L-def}, such that $\delta \dot{\vec{L}} = \mu \; \vec{r}_{12} \times \delta \vec{a}$, where $\delta \vec{a}$ is the non-GR part of Eq.~\eqref{eq:rel-a}.
Evaluating the forces in Eq.~\eqref{eq:f-dCS}-\eqref{eq:f-MQ}, the end result is $\delta \dot{\vec{L}} = \delta \dot{\vec{L}}_{\dd} + \delta \dot{\vec{L}}_{\MQ}$ with
\begin{align}
\label{eq:Ldot-DD}
\delta \dot{\vec{L}}_{\dd} &= - \frac{75}{16} \frac{\zeta_{12}}{r_{12}^{3}} \left[\left(\vec{S}_{1} \cdot \vec{n}_{12}\right) \left(\vec{n}_{12} \times \vec{S}_{2}\right) 
\right.
\nn \\
&\left.
+ \left(\vec{S}_{2} \cdot \vec{n}_{12}\right) \left(\vec{n}_{12} \times \vec{S}_{1}\right)\right]\,,
\\
\label{eq:Ldot-MQ}
\delta \dot{\vec{L}}_{\MQ} &= \frac{603}{112} \frac{1}{r_{12}^{3}} \left[\zeta_{1} \frac{m_{2}}{m_{1}} \left(\vec{S}_{1} \cdot \vec{n}_{12}\right) \left(\vec{n}_{12} \times \vec{S}_{1}\right) 
\right.
\nn \\
&\left.
+ \zeta_{2} \frac{m_{1}}{m_{2}} \left(\vec{S}_{2} \cdot \vec{n}_{12}\right) \left(\vec{n}_{12} \times \vec{S}_{2}\right)\right]\,,
\end{align}
where we have combined the contributions from $\Delta F^{j}_{\Gamma, {\dcs}}$ and $\Delta F^{j}_{\MQ}$ into $\delta \dot{\vec{L}}_{\MQ}$. Both of these terms represent monopole-quadrupole interactions: $\Delta F^{j}_{\Gamma, {\dcs}}$ describes the interaction of the mass monopole of one body with the quadrupole correction to the Newtonian potential of the other body, while $\Delta F^{j}_{\MQ}$ describes the interaction of the mass quadrupole of one body with the monopole Newtonian potential of the other body. This completes the derivation of the angular momentum evolution in dCS gravity.

%%%%%%%%%%%%%%%%%%%%%%%%%%%%%%%%%%%
\section{Properties of the PN Precession Equations}
\label{na}

In this section, we discuss the existence of conserved quantities in the spin dynamics necessary for the construction of analytic Fourier domain waveforms for spin-precessing BH binaries in dCS gravity, and then present some general properties of precession of the waveforms. 

%--------------------------------------------------------------
\subsection{Conserved Quantities for Spin-Precessing Binaries in dCS Gravity} 
The problem of constructing analytic waveforms for generic spin-precessing BH binaries in GR is one that has only recently been solved. The breakthrough that allowed this was the development of an analytic, closed form solution to the leading PN order spin-precession equations in GR~\cite{Racine:2008qv, Kesden:2014sla}. This was extended to include the effect of radiation reaction on the binary using multiple scale analysis (MSA)~\cite{Chatziioannou:2017tdw}. Finally, Fourier domain waveforms were computed in the stationary phase approximation through the application of shifted uniform asymptotics (SUA) in~\cite{Chatziioannou:2017tdw, Chatziioannou:2016ezg}. The latter of these two are generic mathematical techniques and will apply in any theory of gravity. However, the development of an analytic solution to the spin-precession equations in GR requires the existence of certain conserved integrals of motion. We here consider the conservation of these quantities in dCS gravity.

The GR precession equations admit seven conserved quantities, namely the magnitudes of the spins and orbital angular momentum $(S_{1}, S_{2}, L)$, total angular momentum vector $\vec{J} = \vec{L} + \vec{S}_{1} + \vec{S}_{2}$, and the mass-weighted effective spin $\Xi$, given in GR by
\begin{equation}
\label{eq:Xi-GR}
\Xi_{\gr} = \left(1 + q\right) \left(\vec{S}_{1} \cdot \hat{L}\right) + \left(1 + \frac{1}{q}\right) \left(\vec{S}_{2} \cdot \hat{L}\right)\,,
\end{equation}
where $q = m_{2}/m_{1}$ and $\hat{L}$ is the unit orbital angular momentum vector. First, consider the total angular momentum vector. The existence of a conserved total angular momentum is obvious from symmetry. In fact, a straightforward calculation from Eqs.~\eqref{eq:Sdot-GR}-\eqref{eq:Sdot-MQ} and~\eqref{eq:Ldot-DD}-\eqref{eq:Ldot-MQ}, reveals that $\vec{J}$ is conserved, i.e. $d\vec{J}/dt = 0$. Conservation of the magnitude of the spin angular momenta $S_{1}$ and $S_{2}$ can be proven formally from either Eq.~\eqref{eq:dSdt-dCS} or Eq~\eqref{eq:dSvecdt-dCS}, and this is shown explicitly in Appendix~\ref{p-to-u}.

Proving the conservation of the two remaining quantities is rather tricky with the precession equations we have at hand. Fortunately, the problem can be simplified somewhat by working with the orbit-averaged precession equations. Orbit averaging is only justified if there is a separation of scales in the problem, or more specifically, if the angular momenta change on a timescale that is significantly longer than the orbital period of the binary. This timescale, called the precession timescale, may be approximated by $T_{\rm prec} = S_{1} / |d\vec{S}_{1}/dt|$. A simple analysis of the precession equations reveals that $T_{\rm prec} \sim M v^{-5}$, while the orbital timescale is just the orbital period, $T_{\ORB} \sim M v^{-3}$, where $v^{2} \sim M/r_{12}$ is the orbital velocity of the binary to leading PN order. Comparing these timescales, we have $T_{\ORB}/T_{\rm prec} \sim v^{2}$. Thus, if the orbital velocity of the binary is small, then $T_{\ORB}/T_{\rm prec} \ll 1$ and the angular momenta evolve on a timescale much longer than the orbital timescale. This implies that if we work in the PN approximation, we are well justified in orbit averaging the precession equations.

To orbit average the precession equations, we must first provide a few details on  the parametrization of the orbit. We consider an orbit with an arbitrary orientation in space. Specifically, the orientation of the binary is governed by three angles: the inclination angle $\iota$ (that subtended by the $z$-axis of our coordinate system and the orbital angular momentum), the longitude of the ascending node $\Omega$ (that subtended by the x-axis and the ascending node of the orbit), and the longitude of pericenter $\omega$ (that subtended by the Runge-Lenz vector and the ascending node). In this parametrization, the radial separation unit vector $\vec{n}_{12}$ and the orbital angular momentum unit vector $\hat{L}$ are given by~\cite{PW},
\begin{align}
\label{eq:n12}
\vec{n}_{12} &= \left({\rm cos}\Omega \; {\rm cos}\phi - {\rm cos}\iota \; {\rm sin}\Omega \; {\rm sin}\phi\right) \vec{e}_{x} 
\nn \\
&+ \left({\rm sin}\Omega \; {\rm cos}\phi + {\rm cos}\iota \; {\rm cos}\Omega \; {\rm sin}\phi\right) \vec{e}_{y} 
\nn \\
&+ {\rm sin}\iota \; {\rm sin}\phi \; \vec{e}_{z}\,,
\\
\label{eq:L-hat}
\hat{L} &= {\rm sin}\iota \; {\rm sin}\Omega \; \vec{e}_{x}  - {\rm sin}\iota \; {\rm cos}\Omega \; \vec{e}_{y}  + {\rm cos}\iota \; \vec{e}_{z}\,,
\end{align}
where $\phi$ and $r_{12}$ are the orbital phase and the orbital separation or orbital radius of the binary. These quantities can be parameterized in terms of orbital elements  via
\begin{align}
r_{12} &= \frac{p}{1 + e \; {\rm cos}\left(\phi - \omega\right)}\,,
\\
\label{eq:phi-dot}
\dot{\phi} &= \left(\frac{m}{p^{3}}\right)^{1/2} \left[1 + e \; {\rm cos}(\phi - \omega)\right]^{2}\,,
\end{align}
where $e$ is the orbital eccentricity and $p$ is the semi-latus rectum of the orbit, and recall that $m$ is the total mass. 

We define orbit averaging in the usual way, through
\be
\left< A \right> := \frac{1}{T_{\ORB}} \int_{0}^{T_{\ORB}} A \; dt\,,
\ee
for any quantity $A$. The orbit average of the GR precession equations has already been studied in detail in the literature, so we will focus on the dCS modifications. In order to obtain the average of these, we must compute the average of $n_{12}^{i} n_{12}^{j} / r_{12}^{3}$. Applying Eqs.~\eqref{eq:n12}-\eqref{eq:L-hat}, we find
\begin{align}
\Bigg\langle \frac{n_{12}^{i} n_{12}^{j}}{r_{12}^{3}} \Bigg\rangle &= \frac{\left(1 - e^{2}\right)^{3/2}}{2 p^{3}}\left(\delta^{ij} - \hat{L}^{i} \hat{L}^{j}\right)
\end{align}
where $\delta^{ij}$ is the Kronecker delta. Thus, the orbit-averaged precession equations in dCS gravity become,
\allowdisplaybreaks[4]
\begin{widetext}
\begin{align}
\label{eq:S1dot-avg-dCS}
\Bigg\langle \frac{d\vec{S}_{1}}{dt} \Bigg\rangle &= \left(1 - e^{2}\right)^{3/2} \left[ \frac{\eta M^{3/2}}{p^{5/2}} \left(2 + \frac{3}{2} q\right) - \frac{3}{2 p^{3}} \left(1 + \frac{25}{16} \zeta_{12}\right)  \left(\hat{L} \cdot \vec{S}_{2}\right) - \frac{3}{2} \frac{q}{p^{3}} \left(1 - \frac{201}{112}\zeta_{1}\right) \left(\hat{L} \cdot \vec{S}_{1} \right) \right] \hat{L} \times \vec{S}_{1} 
\nn \\
&+ \frac{(1 - e^{2})^{3/2}}{2 p^{3}} \left(1 + \frac{25}{16} \zeta_{12}\right) \vec{S}_{2} \times \vec{S}_{1}\,,
\\
\label{eq:S2dot-avg-dCS}
\Bigg\langle \frac{d\vec{S}_{2}}{dt} \Bigg\rangle &= \left(1 - e^{2}\right)^{3/2} \left[ \frac{\eta M^{3/2}}{p^{5/2}} \left(2 + \frac{3}{2q}\right) - \frac{3}{2 p^{3}} \left(1 + \frac{25}{16} \zeta_{12}\right)  \left(\hat{L} \cdot \vec{S}_{1}\right) - \frac{3}{2} \frac{1}{q p^{3}} \left(1 - \frac{201}{112} \zeta_{2}\right) \left(\hat{L} \cdot \vec{S}_{2} \right) \right] \hat{L} \times \vec{S}_{2} 
\nn \\
&+ \frac{(1 - e^{2})^{3/2}}{2 p^{3}} \left(1 + \frac{25}{16} \zeta_{12}\right) \vec{S}_{1} \times \vec{S}_{2}\,,
\\
\label{eq:Ldot-avg-dCS}
\Bigg\langle \frac{d\hat{L}}{dt} \Bigg\rangle &= \left(1 - e^{2}\right)^{3/2} \left[ \frac{1}{p^{3}} \left(2 + \frac{3}{2}q\right) - \frac{3}{2 p^{7/2}} \left(1 + \frac{25}{16} \zeta_{12}\right)  \frac{\left(\hat{L} \cdot \vec{S}_{2}\right)}{\eta M^{3/2}} - \frac{3}{2} \frac{q}{p^{7/2}} \left(1 - \frac{201}{112} \zeta_{1}\right) \frac{\left(\hat{L} \cdot \vec{S}_{1} \right)}{\eta M^{3/2}} \right] \vec{S}_{1} \times \hat{L}
\nn \\
&+\left(1 - e^{2}\right)^{3/2} \left[ \frac{1}{p^{3}} \left(2 + \frac{3}{2q}\right) - \frac{3}{2 p^{7/2}} \left(1 + \frac{25}{16} \zeta_{12}\right)  \frac{\left(\hat{L} \cdot \vec{S}_{1}\right)}{\eta M^{3/2}} - \frac{3}{2} \frac{1}{q p^{7/2}} \left(1 - \frac{201}{112} \zeta_{2}\right) \frac{\left(\hat{L} \cdot \vec{S}_{2} \right)}{\eta M^{3/2}} \right] \vec{S}_{2} \times \hat{L}\,,
\end{align}
\end{widetext}
where we have used that $L = \eta M^{3/2} p^{1/2}$ is the magnitude of the orbital angular momentum at leading PN order.

A quick study of these equations reveals that $(\vec{S}_{1}, \vec{S}_{2}, \hat{L})$ only change according to the cross products $(\vec{S}_{1} \times \hat{L}, \vec{S}_{2} \times \hat{L}, \vec{S}_{1} \times \vec{S}_{2})$, implying that any changes to the spin and angular momentum vectors are orthogonal to their directions. As a result, the magnitude of these vectors are conserved under the influence of precession. Note that this is, however, not strictly true, since the binary will inspiral due to the emission of GWs. At leading PN order, the binary loses orbital angular momentum due to radiation reaction, and as a result, the magnitudes $L = |\vec{L}|$ and $J = |\vec{J}|$ are no longer conserved. Further, GWs can also be lost through the horizon of the BHs, causing changes to $|\vec{S}_{1}|$ and $|\vec{S}_{2}|$ through tidal heating and tidal torquing~\cite{Chatziioannou:2012gq, Chatziioannou:2016kem}, although this effect is much higher PN order. Regardless, we are here only concerned with the conservative dynamics of the binary under the influence of precession, and we shall not consider the full effects of radiation reaction in dCS gravity.

We have thus far proven that six quantities that are conserved in GR, $(S_{1},S_{2},L,\vec{J})$, are still conserved in dCS gravity, but what about the mass-weighted effective spin $\Xi$? A straightforward computation, by taking a time derivative of Eq.~\eqref{eq:Xi-GR} and applying the dCS precession equations in Eqs.~\eqref{eq:S1dot-avg-dCS}-\eqref{eq:Ldot-avg-dCS}, reveals that the GR mass-weighted effective spin $\Xi_{\gr}$ is no longer conserved, and that its non-conservation is linear in the dCS coupling constants $(\zeta_{1}, \zeta_{2}, \zeta_{12})$. The question is then whether a new mass-weighted effective spin can be constructed in dCS gravity that is conserved under the dCS precession equations. To do this, let us use the ansatz
\begin{align}
\Xi_{\dcs} &= (1 + q) \left[1 + \zeta_{1} {\cal{A}}\left(\hat{L} \cdot \vec{S}_{1}, \hat{L} \cdot \vec{S}_{2}\right) 
\right.
\nn \\
&\left.
+ \zeta_{12} {\cal{C}}\left(\hat{L} \cdot \vec{S}_{1}, \hat{L} \cdot \vec{S}_{2}\right)\right] \hat{L} \cdot \vec{S}_{1}
\nn \\
&+ \left(1 + \frac{1}{q}\right) \left[1 + \zeta_{2} {\cal{B}}\left(\hat{L} \cdot \vec{S}_{1}, \hat{L} \cdot \vec{S}_{2}\right) 
\right.
\nn \\
&\left.
+ \zeta_{12} {\cal{C}}\left(\hat{L} \cdot \vec{S}_{1}, \hat{L} \cdot \vec{S}_{2}\right)\right] \hat{L} \cdot \vec{S}_{2}
\end{align}
with undetermined functions $({\cal{A}},{\cal{B}},{\cal{C}})$ and require $d\Xi_{\dcs}/dt = 0$. Linearizing in $(\zeta_{1}, \zeta_{2}, \zeta_{12})$, we obtain differential equations for the functions $({\cal{A}}, {\cal{B}}, {\cal{C}})$, which may be solved to obtain
\begin{align}
{\cal{A}}\left({\cal{Y}}_{1}, {\cal{Y}}_{2}\right) &= \frac{201}{224} \frac{q {\cal{Y}}_{1}}{(1 + q) L - q {\cal{Y}}_{1} - {\cal{Y}}_{2}}\,,
\\
{\cal{B}}\left({\cal{Y}}_{1}, {\cal{Y}}_{2}\right) &= -\frac{201}{224} \frac{q {\cal{Y}}_{1} \left(q {\cal{Y}}_{1} + 2 {\cal{Y}}_{2}\right)}{{\cal{Y}}_{2} \left[(1 + q) L - q {\cal{Y}}_{1} - {\cal{Y}}_{2}\right]}\,,
\\
{\cal{C}}\left({\cal{Y}}_{1}, {\cal{Y}}_{2}\right) &= \frac{25}{48} \frac{q {\cal{Y}}_{1} \left[(q - 1) L - 3 {\cal{Y}}_{2}\right]}{\left(q {\cal{Y}}_{1} + {\cal{Y}}_{2}\right) \left[(1 + q) L - q {\cal{Y}}_{1} - {\cal{Y}}_{2}\right]}\,,
\end{align}
where ${\cal{Y}}_{1} = \hat{L} \cdot \vec{S}_{1}$ and ${\cal{Y}}_{2} = \hat{L} \cdot \vec{S}_{2}$. Thus, we have shown that there exists a conserved mass-weighted effective spin in dCS gravity to linear order in the coupling. This completes the discussion of the constants of motion of the precession problem in dCS gravity.

%--------------------------------------------------------------
\subsection{The Effect of Precession on Gravitational Waves in dCS Gravity} 
As a final point, we illustrate the differences induced by precession in dCS gravity compared to GR. To illustrate this, we numerically solve the precession equations using \texttt{Mathematica}, for a circular binary system with $(m_{1}, m_{2}) = (10,10) M_{\odot}$ and $(\chi_{1}, \chi_{2}) = (0.9, 0.95)$. We choose the spins to be oriented such that $\hat{S}_{1} = (0.9, 0, 0.44)$ and $\hat{S}_{2} = (0.1, 0.25, 0.96)$, and choose $\hat{J} = (0,0,1)$, which fixes the direction of the orbital angular momentum. We evolve the binary under leading PN order radiation reaction in GR, which forces the semi-latus rectum of the orbit to evolve via
\begin{equation}
\label{eq:dpdt}
\frac{dp}{dt} = - \frac{64}{5} \eta \left(\frac{M}{p}\right)^{3}
\end{equation}
for circular orbits. We start the evolution such that the GW frequency, which is specifically $f_{\rm GW} = 2/T_{\ORB}$, is 10 Hz, and evolve the binary up to the last stable orbit at $p = r_{12} = 6 M$. We use a polarization basis such that the plus and cross polarizations are given by~\cite{Apostolatos:1994mx}
\begin{align}
\label{eq:hplus}
h_{+} &= - \frac{2 \eta M^{2}}{p D_{L}} \left[1 + \left(\hat{L} \cdot \vec{N}\right)^{2} \right] \; {\rm cos}(2 \phi)
\\
\label{eq:hcross}
h_{\times} &= - \frac{4 \eta M^{2}}{p D_{L}} \left(\hat{L} \cdot \vec{N}\right) \; {\rm sin}(2 \phi)
\end{align}
where $\vec{N} = [{\rm sin}\Theta \; {\rm cos}\Phi, {\rm sin}\Theta \; {\rm cos}\Phi, {\rm cos}\Theta]$ is the direction to the source on the sky, and $\phi$ is the orbital phase of the binary, which is governed by $d\phi/dt = (M/p)^{3/2}$. We consider two cases, one in which the waveforms are computed purely in GR (i.e.~$\xi = 0$), and one in dCS gravity with $\xi^{1/4} = 10 \; {\rm km}$, which corresponds to $\zeta_{1} = \zeta_{2} = \zeta_{12} \sim 0.2$. Further, we take $\Theta = \pi/4$ and $\Phi = 0$.

\begin{figure}[ht]
\includegraphics[clip=true,width=\columnwidth]{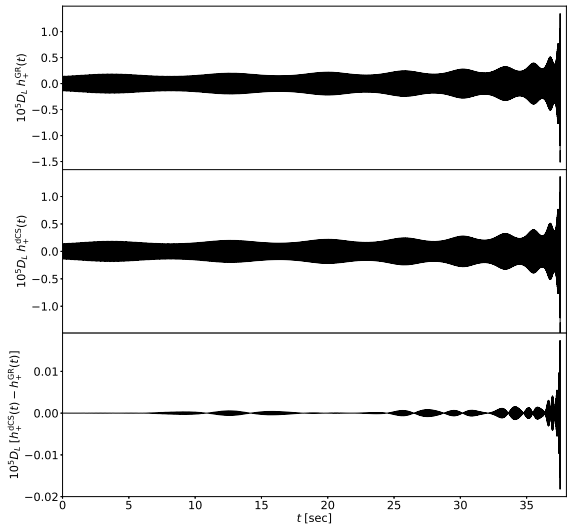}
\caption{\label{wave} Comparison of the leading PN order waveform of a spin-precessing binary BH with $m_{1} = m_{2} = 10 M_{\odot}$, $\chi_{1} = 0.9$, and $\chi_{2} = 0.95$. Radiation reaction is included via the leading PN order contribution to $dp/dt$. The top panel displays the waveform with the GR precession effects $(\xi = 0)$, while the middle panel includes the dCS precession modifications $(\xi^{1/4} = 10 \; {\rm km})$. The bottom panel displays the difference in the two waveforms.}
\end{figure}

The results of this numerical calculation are displayed in Fig.~\ref{wave} for the plus polarization given by Eq.~\eqref{eq:hplus}. In the top panel, we display the waveform with only the precessional effects in GR, while in the middle panel we display the waveform with the dCS precession modifications. The two waveforms display the amplitude modulations present in precessing systems. However, since there are additional precession effects in dCS gravity, the modulation is not identical to GR, and the difference between the two waveforms is displayed in the bottom panel of Fig.~\ref{wave}. 

Clearly, these waveforms are not realistic representations of the GWs from this binary in dCS gravity, since we are assuming that the radiation reaction force and gravitational waveform are still governed by the leading PN order GR effects given in Eq.~\eqref{eq:dpdt}-\eqref{eq:hcross}. In dCS gravity, inspiraling binary BHs also emit scalar dipole radiation, and the waveform picks up a 2PN correction proportional to the scalar dipole moments of the BHs. We are here only interested in how the specific precession effects are different from GR, and we leave the inclusion of these radiation reaction effects to future work.

%

%%%%%%%%%%%%%%%%%%%%%%%%%%%%%%%%%%%
\section{Discussion}
\label{conclusion}

We have derived the leading PN order spin-precession equations for binary BHs in dCS gravity. We have constructed an EFT description for BHs in dCS gravity, which is only characterized by two undetermined coefficients, specifically $C_{\vartheta}^{\dcs}$ and $\delta C_{\QUAD}^{\dcs}$. The first of these controls the interaction of the scalar dipole moment of the BH with an external pseudo-scalar field $\vartheta$. The second is the correction to the BH's quadrupole moment $C_{\QUAD}$ due to the modified spin induced quadrupole in dCS gravity. We were able to fix the values of both of these constants by solving the field equations in a NZ expansion, and matching the solutions to the isolated BH spacetimes found in~\cite{kent-CSBH} for example. While this is not the first application of EFT methods to modified theories of gravity~\cite{Sennett:2017lcx}, it is the first application to spin-precessing binaries in these theories. In fact, our analysis is the first to consider how BH binaries precess in any modified theory of gravity. 

With the NZ field solutions in hand, we proceeded to expand the precession equations in the weak field, slow motion approximation. We found that the precession equations pick up two modification from GR, a scalar dipole-dipole interaction which takes the same form as the leading PN order spin-spin interaction, and a modified quadrupole-monopole interaction. We have shown numerically that these extra modifications affect the amplitude modulation of GWs emitted by inspiraling and precessing binary BHs. Further, we have analytically verified the existence of seven constants of motion in the precession problem when neglecting radiation reaction. The existence of these seven constants should allow for the construction of an analytic, closed form solution to the precession equations in dCS gravity. Once such a solution is constructed, one may then include radiation reaction by promoting the appropriate constants of motion, such as the orbital angular momentum $L$, to be functions of time that vary slowly over the orbital and precession timescales. This method works provided there is a separation of scales between orbital, precession, and radiation-reaction timescales, which is well defined in the PN approximation. This will lead the way to the development of analytic Fourier domain waveforms for spin-precessing binary BHs in dCS gravity along the lines of those that have been constructed in GR.

With these waveforms in hand, one may consider how well dCS gravity may be constrained with ground-based and space-based GW detectors, similar to what has been done using the currently detected GW signals~\cite{Yunes:2016jcc, TheLIGOScientific:2016src}. One could also postulate the existence of new ppE~\cite{PPE} parameters associated specifically with modification to the precessional dynamics of binary systems. We have here shown that two modifications appear in dCS gravity, and we may expect that these can be described by two ppE parameters in a Fourier domain waveform. These two parameters only modify the spin-spin and quadrupole-monopole interaction, as can be seen in Eqs.~\eqref{eq:S1dot-avg-dCS}-\eqref{eq:Ldot-avg-dCS}. In dCS gravity, the spin-orbit interaction is not modified, but this may not be true in other modifies theories. One theory that may produce such a modification is Einstein-dilaton-Gauss-Bonnet (EDGB) gravity, the even-parity ``sister'' theory to dCS gravity. In EDGB gravity, BHs possess a scalar monopole charge. Just like an electron-positron pair produces a magnetic monopole moment in electromagnetism, BHs in EDGB likely generate a magnetic type scalar dipole moment associated with the orbital motion of the binary, and may modify the spin-orbit interaction. In this way, we may obtain a set of ppE parameters that describe the modifications to the leading PN order spin-orbit, spin-spin, and quadrupole-monopole interactions. The work we have completed here has moved us one step closer to understanding how to probe fundamental physics with the GW emission from spin-precessing binary systems.

%%%%%%%%%%%%%%%%%%%%%%%%%%%%%%%%%%%
\acknowledgements

N. L. and N. Y. acknowledge support from NSF EAPSI Award No. 1614203, NSF CAREER grant PHY-1250636, and NASA grants NNX16AB98G and 80NSSC17M0041. T. T. acknowledges support in part by MEXT Grant-in-Aid for Scientific Research on Innovative Areas, Nos. 17H06357 and 17H06358, and by Grant-in-Aid for Scientific Research Nos. 26287044 and 15H02087. We would like to thank Chad Galley for providing a useful set of notes on EFT to aid in this work.
%%%%%%%%%%%%%%%%%%%%%%%%%%%%%%%%%%%%%%%%%%%%
\appendix
%%%%%%%%%%%%%%%%%%%%%%%%%%%%%%%%%%%
\section{Parity Transformations}
\label{parity}

Typically, one cannot assign absolute parity to spacetime tensors, since these quantities contain both spatial and temporal components. To determine the parity of such objects, we may foliate spacetime with a set of suitable chosen space-like hypersurfaces, and perform parity transformations on these hypersurfaces~\cite{Alexander:2017jmt}. Clearly then, this concept of parity is foliation-dependent. However, in this paper, we wish to assign parity to spacetime tensors, namely to the tetrad $e^{\mu}_{A}$ and the Levi-Civita tensor $\epsilon_{\mu \nu \rho \sigma}$, which transform under parity through Eqs.~\eqref{eq:parity-def}. This appendix provides a review of the usual way of doing parity transformations, and shows that one obtains the same result as when using our definitions.

To begin, it is useful to consider parity transformations in classical Newtonian mechanics. Vectors in Newtonian mechanics are three-dimensional spatial quantities, defined generically as
\begin{equation}
\label{eq:vec-A}
\vec{A} = A^{i} \vec{e}_{i}\,,
\end{equation}
where $\vec{e}_{i}$ are the coordinate basis vectors and $A^{i}$ are scalar quantities called vector components. Parity transformations are complete spatial inversions which force basis vectors to point in the opposite direction relative to their initial direction, specifically $\hat{P}[\vec{e}_{i}] = - \vec{e}_{i}$. These quantities are said to be odd under a parity transformation. Scalar quantities, such as the individual components of a vector, are even under parity, specifically $\hat{P}[A^{i}] = + A^{i}$, which then implies that vectors are odd under a parity transformation. Such a view of transformations is sometimes referred to as \emph{active}. 

Let us now consider an object constructed from the three dimensional cross product of two vectors. The orbital angular momentum of a particle, for example, is given by
\begin{equation}
\vec{L} = \vec{x} \times \vec{p} = x^{i} p^{j} \vec{e}_{i} \times \vec{e}_{j}\,,
\end{equation}
where $\vec{x}$ and $\vec{p}$ are the position and momentum vectors of the particle, respectively. The cross product of the two basis vectors $\vec{e}_{i} \times \vec{e}_{j}$ is even under a parity transformation, and thus, so is the orbital angular momentum. We thus recognize $\vec{L}$ as a pseudovector. The cross product can, in general, be used to define the three dimensional Levi-Civita symbols, a collection of scalars that must be even under parity if $\vec{L}$ is to also be even under parity. By following the above procedures, we may assign definitive parity to any object in Newtonian mechanics.

Now, let us consider parity transformations on a spacetime manifold. To do this, we must specify the hypersurfaces on which we will perform the parity transformations. Spacetime coordinates will be specified by $x^{\mu}$, while coordinates on the tangent space are $y^{A}$. For the problem considered here, it is convenient to choose the normal to $y^{0} = \text{const.}$ hypersurfaces to be the particle's four velocity $u^{\mu}$ and to perform the projections onto this hypersurface using the tetrad 
\begin{equation}
\bar{e}^{\mu}_{A} = \frac{\partial x^{\mu}}{\partial y^{A}}\,,
\end{equation}
where we have placed a bar on this tetrad to distinguish it from the co-rotating tetrad used in the main text of this paper. Our choice for the normal vector forces $\bar{e}^{\mu}_{0} = u^{\mu}$. The parity transformation is performed on the hypersurface, specifically $y^{I} \rightarrow - y^{I}$, where $I$ indicates the purely spatial components of $y^{A}$. This implies that $\hat{P}[\bar{e}^{\mu}_{I}] \rightarrow - \bar{e}^{\mu}_{I}$, which are now analogous to the basis vectors used in Newtonian mechanics.

In the main text of this paper, we considered parity transformations when considering possible terms that could be added to the effective matter Lagrangian to describe the scalar dipole moment of BHs in dCS gravity. We thus focus on proving that the scalar dipole term, the third term in Eq.~\eqref{eq:L-dCS}, is even under parity. Just as in the main text, we do not distinguish between the direction of ${\cal{P}}^{\mu}$ and that of $u^{\mu}$, thus neglecting higher-order-in-spin terms. Our choice of SSC implies that $S_{\mu} u^{\mu} = 0$ to the order we are working. Thus, $S_{I} = S_{\mu} \bar{e}^{\mu}_{I}$ is the spatial spin vector. This spin vector is the usual one that appears in Newtonian mechanics, which is there defined as the cross product between the dipole moment of a fluid element and its 3-velocity. Therefore, the spin, like the angular momentum, is actually a pseudo-vector, specifically $\hat{P}[S_{I}] \rightarrow + S_{I}$. If we now apply the hypersurface projections on the scalar dipole term in Eq.~\eqref{eq:L-dCS}, we have
\begin{align}
\frac{1}{2} \epsilon_{\mu \nu \rho \sigma} u^{\mu} S^{\rho \sigma} \nabla^{\nu} \vartheta &= S^{\nu} \nabla_{\nu} \vartheta = S^{I} \nabla_{I} \vartheta\,.
\end{align}
The dCS scalar field is odd under parity and its spatial derivative on this hypersurface is even $\hat{P}[\nabla_{I} \vartheta] \rightarrow + \nabla_{I} \vartheta$. Thus, the total scalar dipole term in Eq.~\eqref{eq:L-dCS} is even under parity and the action is invariant under a parity transformation.

Now, consider the parity transformations given by Eq.~\eqref{eq:parity-def}. The parity of the co-rotating tetrad implies that the rotation tensor $\Omega^{\mu \nu}$ has even parity by Eq.~\eqref{eq:Omega-def}. Since the spin tensor is defined as the momentum associated with $\Omega^{\mu \nu}$, it must also be even under parity. Further, since we have only assigned odd parity to $e^{\mu}_{I}$ and $\epsilon_{\mu \nu \rho \sigma}$, the four velocity of the particle $u^{\mu} = d z^{\mu}/d\tau = e^{\mu}_{I} (D z^{I}/D\tau)$ must be even under parity because $\hat{P} \left[z^{I}\right] = - z^{I}$ so that $\hat{P} \left[z^{\mu}\right] = + z^{\mu}$, while $\hat{P}[\nabla_{\alpha} \vartheta] \rightarrow - \nabla_{\alpha} \vartheta$ since $\nabla_{\alpha} \vartheta = e_{\alpha}^{I} \nabla_{I} \vartheta$. Applying the parity transformation to the scalar dipole term in Eq.~\eqref{eq:L-dCS}, we find
\begin{equation}
\hat{P}\left[\epsilon_{\mu \nu \rho \sigma} u^{\mu} S^{\rho \sigma} \nabla^{\nu} \vartheta\right] \rightarrow + \epsilon_{\mu \nu \rho \sigma} u^{\mu} S^{\rho \sigma} \nabla^{\nu} \vartheta\,,
\end{equation}
thus recovering the behavior from the hypersurface parity transformation.
%%%%%%%%%%%%%%%%%%%%%%%%%%%%%%%%%%%
\section{Properties of the Particle Equations of Motion in the EFT Formalism}
\label{p-to-u}

In the main text of this paper, we considered the properties of the PN expanded precession equations given in Eqs.~\eqref{eq:S1dot-avg-dCS}-\eqref{eq:Ldot-avg-dCS}. We here provide a description of the properties of the canonical equations of motion derived from EFT and given in Eqs.~\eqref{eq:dpdt-dCS}-\eqref{eq:dSdt-dCS}, specifically focusing on the relationship between the particle's canonical momentum and the four velocity, as well as constants of motion along the particle's worldline.

We begin by deriving the relationship between the canonical momentum ${\cal{P}}_{\mu}$ and the four velocity $u^{\alpha}$ for our EFT given by the Lagrangian in Eq.~\eqref{eq:L-dCS}. Rather than restricting attention to the choice of SSC given in the main text $S^{\mu \nu} {\cal{P}}_{\nu} = 0$, we consider the more general SSC
\begin{equation}
S^{\mu \nu} f_{\nu} = 0\,, \qquad S^{\mu} f_{\mu} = 0\,,
\end{equation}
where $f_{\mu}$ is a time-like vector. Taking the derivative of the SSC, we have
\begin{equation}
\label{eq:SSC-dt}
\frac{DS^{\mu \nu}}{D\tau} f_{\nu} = - S^{\mu \nu} \frac{Df_{\nu}}{D\tau}\,.
\end{equation}
Consider the left-hand side of this equation. Applying the precession equation in Eq.~\eqref{eq:dSdt-dCS}, we have
\begin{align}
\frac{DS^{\mu \nu}}{D\tau} f_{\nu} &= (u^{\nu} f_{\nu}) {\cal{P}}^{\mu} - ({\cal{P}}^{\nu} f_{\nu}) u^{\mu}  
\nn \\
&+ \frac{2}{m^{2}} C_{\vartheta}^{\dcs} {^{\star} S_{\alpha}}^{[\mu} f_{\nu} \nabla^{\nu]} \vartheta u^{\alpha}
\nn \\
&+ \frac{4}{3} {R_{\alpha \beta \rho}}^{[\mu} f_{\nu} J^{\nu] \rho \beta \alpha}\,.
\end{align}
Inserting this back into Eq.~\eqref{eq:SSC-dt} and solving for ${\cal{P}}^{\mu}$, we find
\begin{align}
{\cal{P}}^{\mu} &= \frac{1}{-f_{\alpha} u^{\alpha}} \left[\left(- f_{\nu} {\cal{P}}^{\nu}\right) u^{\mu} + S^{\mu \nu} \frac{D f_{\nu}}{D\tau} 
\right.
\nn \\
&\left.
+ \frac{2}{m^{2}} C_{\vartheta}^{\dcs} {^{\star} S_{\alpha}}^{[\mu} f_{\nu} \nabla^{\nu]} \vartheta u^{\alpha} + \frac{4}{3} {R_{\alpha \beta \rho}}^{[\mu} f_{\nu} J^{\nu] \rho \beta \alpha}\right]\,,
\end{align}
which is the generalized expression for ${\cal{P}}^{\mu}(u^{\alpha})$.

Let us now define the mass and spin of the particle via
\begin{align}
\label{eq:mass-spin-def}
{\cal{M}}^{2} = - {\cal{P}}_{\alpha} {\cal{P}}^{\alpha}\,,
\qquad
S^{2} = \frac{1}{2} S_{\mu \nu} S^{\mu \nu}\,.
\end{align}
and restrict attention to the case $f_{\mu} = {\cal{P}}_{\mu}$. We then find that
\begin{align}
{\cal{P}}^{\mu} &= \frac{1}{-{\cal{P}}_{\alpha} u^{\alpha}} \left[ {\cal{M}}^{2} u^{\mu} + \frac{1}{2} S^{\mu \nu} S_{\beta \gamma} {R^{\beta \gamma}}_{\rho \nu} u^{\rho} 
\right.
\nn \\
&\left.
- \frac{1}{6} S^{\mu \nu} J^{\beta \gamma \delta \rho} \nabla_{\nu} R_{\beta \gamma \delta \rho} + \frac{4}{3} {R_{\alpha \beta \rho}}^{[\mu} {\cal{P}}_{\nu} J^{\nu] \rho \beta \alpha} 
\right.
\nn \\
&\left.
+ \frac{1}{m^{2}} C_{\vartheta}^{\dcs} S^{\mu \nu} {^{\star} S_{\gamma}}^{\beta} u^{\gamma} \nabla_{\nu \beta} \vartheta 
\right.
\nn \\
&\left.
+ \frac{2}{m^{2}} C_{\vartheta}^{\dcs} {^{\star} S_{\alpha}}^{[\mu} {\cal{P}}_{\nu} \nabla^{\nu]} \vartheta u^{\alpha}\right]\,.
\end{align}
This expression can be simplified by working in a small-spin, weak-coupling approximation. Recall that $\vartheta \sim {\cal{O}}(S)$ and $J \sim {\cal{O}}(S^{2})$, and that the canonical momentum depends on the bare momentum and dCS scalar field through Eq.~\eqref{eq:can-mom}. Applying this, we have
\begin{align}
\label{eq:P-of-u}
{\cal{P}}^{\mu} &= \frac{1}{- {\cal{P}}_{\alpha} u^{\alpha}} \left[{\cal{M}}^{2} u^{\mu} + \frac{1}{2} S^{\mu \nu} S_{\beta \gamma} {R^{\beta \gamma}}_{\rho \nu} u^{\rho} 
\right.
\nn \\
&\left.
+ \frac{2}{m^{2}} C_{\vartheta}^{\dcs} {^{\star} S_{\alpha}}^{[\mu} p_{\nu} \nabla^{\nu]} \vartheta u^{\alpha} + \frac{4}{3} {R_{\alpha \beta \rho}}^{[\mu} p_{\nu} J^{\nu] \rho \beta \alpha}\right]
\nn \\
&+ {\cal{O}}(S^{3})\,.
\end{align}
To obtain ${\cal{P}}_{\alpha} u^{\alpha}$, one simply has to contract the above equation with $u_{\mu}$, specifically
\begin{align}
\left({\cal{P}}^{\mu} u_{\mu}\right)^{2} &= - {\cal{M}}^{2} u^{2} - \frac{1}{2} u_{\mu} S^{\mu \nu} S_{\beta \gamma} {R^{\beta \gamma}}_{\rho \nu} u^{\rho} 
\nn \\
&+ \frac{1}{m^{2}} C_{\vartheta}^{\dcs} u_{\mu} {^{\star} S_{\alpha}}^{\nu} p_{\nu} \nabla^{\mu} \vartheta u^{\alpha}
\nn \\
&- \frac{4}{3} u_{\mu} {R_{\alpha \beta \rho}}^{[\mu} p_{\nu} J^{\nu] \rho \beta \alpha}\,.
\end{align}
This completes the relationship between the canonical momentum and the four velocity.

Finally, we consider the possibility of constants of motion along the particle's worldline, specifically the mass and spin of the particle, which recall were defined in Eq.~\eqref{eq:mass-spin-def}. First, we consider the conservation of the spin of the particle. Taking a time derivative of the above definitions gives
\begin{align}
\frac{D S^{2}}{D\tau} &= S^{\mu \nu} \frac{D S_{\mu \nu}}{D\tau}
\\
&= 2 S^{\mu \nu} {\cal{P}}_{\mu} u_{\nu} + \frac{2}{m^{2}} C_{\vartheta}^{\dcs} S^{\mu \nu} \; {^{\star} S}_{\alpha \mu} \nabla_{\nu} \vartheta \; u^{\alpha}
\nn \\
&+ \frac{4}{3} S^{\mu \nu} R_{\alpha \beta \rho \mu} {J_{\nu}}^{\rho \beta \alpha}
\end{align}
where we have used Eq.~\eqref{eq:dSdt-dCS} in the second equality. The first of these terms vanishes by our choice of SSC. For the remaining two terms, recall that we work in a small spin approximation. Thus, the second term is actually ${\cal{O}}(S^{3})$ since the scalar field in linear in spin, and the third term is also ${\cal{O}}(S^{3})$. As a result, the spin of the particle is conserved to the order considered here.

Now, consider the conservation of the particle's mass. Following the same procedure as the spin, we take a time derivative of Eq.~\eqref{eq:mass-spin-def} and use Eq.~\eqref{eq:dpdt-dCS} to obtain
\begin{align}
\label{eq:dmdt}
\frac{D{\cal{M}}^{2}}{D\tau} &= S_{\alpha \beta} {R^{\alpha \beta}}_{\rho \mu} u^{\rho} {\cal{P}}^{\mu} + \frac{2}{m^{2}} C_{\vartheta}^{\dcs} {^{\star} S_{\alpha}}^{\beta} u^{\alpha} {\cal{P}}^{\mu} \nabla_{\mu \beta} \vartheta 
\nn \\
&- \frac{1}{3} J^{\alpha \beta \gamma \delta} {\cal{P}}^{\mu} \nabla_{\mu} R_{\alpha \beta \gamma \delta}\,.
\end{align}
To simplify the first term, we use the relationship between ${\cal{P}}_{\mu}$ and $u_{\mu}$ given in Eq.~\eqref{eq:P-of-u}, and we find
\begin{align}
S_{\alpha \beta} {R^{\alpha \beta}}_{\rho \mu} &u^{\rho} {\cal{P}}^{\mu} = \frac{{\cal{M}}^{2}}{\left(-{\cal{P}}_{\sigma} u^{\sigma}\right)} S_{\alpha \beta} {R^{\alpha \beta}}_{\rho \mu} u^{\rho} u^{\mu}
\nn \\
& + \frac{1}{2 \left(-{\cal{P}}_{\sigma} u^{\sigma}\right)} S^{\mu \nu} S_{\gamma \delta} {R^{\gamma \delta}}_{\lambda \nu} u^{\lambda} S_{\alpha \beta} {R^{\alpha \beta}}_{\rho \mu} u^{\rho}
\nn \\
&+ {\cal{O}}(S^{3})\,,
\end{align}
where we have applied our small spin expansion on the third and fourth terms in Eq.~\eqref{eq:dpdt-dCS}. The first term above vanishes due to the antisymmetry of the Riemann tensor. One could show that the remaining term is higher order in a small spin expansion, but this term is actually zero exactly. To see this, we rewrite it as
\begin{align}
S^{\mu \nu} S_{\gamma \delta} &{R^{\gamma \delta}}_{\lambda \nu} u^{\lambda} S_{\alpha \beta} {R^{\alpha \beta}}_{\rho \mu} u^{\rho} 
\nn \\
&= S^{\mu \nu} \left(S_{\gamma \delta} {R^{\gamma \delta}}_{\lambda \nu} u^{\lambda}\right) \left(S_{\alpha \beta} {R^{\alpha \beta}}_{\rho \mu} u^{\rho}\right)\,,
\nn \\
&= S^{\mu \nu} A_{\mu} A_{\nu}\,,
\end{align}
where we have defined $A_{\mu} = S_{\alpha \beta} {R^{\alpha \beta}}_{\rho \mu} u^{\rho}$. This vanishes exactly due to the antisymmetry of the spin tensor.

We are now left with simplifying the second and third terms in Eq.~\eqref{eq:dmdt}. Both of these terms follow the same simplification procedure, so we only consider the dipole term. Applying Eq.~\eqref{eq:P-of-u}, we find
\begin{align}
\frac{2}{m^{2}} C_{\vartheta}^{\dcs} {^{\star} S_{\alpha}}^{\beta} u^{\alpha} {\cal{P}}^{\mu} \nabla_{\mu \beta} \vartheta &= \frac{2}{{\cal{M}}} C_{\vartheta}^{\dcs} \; {^{\star} S_{\alpha}}^{\beta} u^{\alpha} \frac{D}{D\tau} \nabla_{\beta} \vartheta
\nn \\
&+ {\cal{O}}(S^{4})
\end{align}
where we have applied a small spin and weak coupling expansion. Now consider,
\begin{align}
{^{\star} S_{\alpha}}^{\beta} u^{\alpha} \frac{D}{D\tau} \nabla_{\beta} \vartheta &= \frac{D}{D\tau} \left({^{\star} S_{\alpha}}^{\beta} u^{\alpha} \nabla_{\beta} \vartheta\right) - \frac{D{^{\star} S_{\alpha}}^{\beta}}{D\tau} u^{\alpha} \nabla_{\beta} \vartheta 
\nn \\
&- {^{\star} S_{\alpha}}^{\beta} \frac{D u^{\alpha}}{D\tau} \nabla_{\beta} \vartheta\,.
\end{align}
We may evaluate the second and third terms in the above equation using Eqs.~\eqref{eq:dpdt-dCS} and~\eqref{eq:dSdt-dCS}. Doing so, we find that these terms are higher order in a small spin expansion. Thus, the overall dipole term in Eq.~\eqref{eq:dmdt} may be converted into a total derivative and moved to the left-hand side. The same procedure may be applied to the quadrupole term in Eq.~\eqref{eq:dmdt}. It then follows that,
\begin{align}
&\frac{D}{D\tau}\left({\cal{M}}^{2} - \frac{2}{{\cal{M}}} C_{\vartheta}^{\dcs} {^{\star} S_{\alpha}}^{\beta} u^{\alpha} \nabla_{\beta} \vartheta 
\right.
\nn \\
&\left.
\;\;\;\;\;\;\;\;\;\;\;\;\;\;\;+ \frac{{\cal{M}}}{3} J^{\alpha \beta \gamma \delta} R_{\alpha \beta \gamma \delta}\right) = {\cal{O}}(S^{4}).
\end{align}
Therefore, it is not the mass ${\cal{M}}$ that is conserved, but the dressed or renormalized mass (the term in parenthesis in the left-hand side of the above equation) that is conserved. This completes the discussion of constants of motion associated with the particle's worldline.
%%%%%%%%%%%%%%%%%%%%%%%%%%%%%%%%%%%
\section{Harmonic Coordinates for Isolated BHs in dCS Gravity}
\label{harmonic}

We here provide the transformation from Boyer-Lindquist coordinates to harmonic coordinates for the isolated BH spacetime in dCS gravity. Harmonics coordinates satisfy the conditions
\begin{equation}
\label{eq:harm-coord}
\Box_{g} X^{(\mu)} = 0\,,
\end{equation}
where $\Box_{g}$ is the D'Alembertian in spacetime with metric $g_{\mu \nu}$, and $X^{(\mu)}$ is a set of scalars that defines harmonic coordinates. Our spacetime is the slowly rotating BH metric of~\cite{kent-CSBH}, which is second order in spin and is written in Boyer-Lindquist coordinate $x^{\mu}_{\BL} = (t, r, \theta, \phi)$. 

A straightforward application of the harmonic coordinate conditions reveals that $t$ and $\phi$ are harmonic, but $r$ and $\theta$ are not. We thus require the mapping from Boyer-Lindquist to harmonic coordinates to only depend on $r$ and $\theta$. We begin by writing
\begin{align}
X^{(t)} :=& t_{h} = t\,,
\\
X^{(x)} :=& x_{h} = r_{h}(r,\theta) {\rm sin}[\theta_{h}(r,\theta)] {\rm cos}\phi\,,
\\
X^{(y)} :=& y_{h} = r_{h}(r,\theta) {\rm sin}[\theta_{h}(r,\theta)] {\rm sin}\phi\,,
\\
X^{(z)} :=& z_{h} = r_{h}(r,\theta) {\rm cos}[\theta_{h}(r,\theta)]\,.
\end{align}
We then solve Eq.~\eqref{eq:harm-coord} in an expansion about $r \gg m$, obtaining
\begin{align}
r_{h}(r, \theta) &= r \left[1 - \frac{m}{r} + \frac{a^{2} {\rm sin}^{2}\theta}{2 r^{2}} - \frac{4 a^{2} m^{2} {\rm sin}^{2} \theta - a^{4} {\rm sin}^{4} \theta}{8 r^{4}} 
\right.
\nn \\
&\left.
+ \frac{25 \bar{\zeta} a^{2} m^{2} (3{\rm cos}^{2}\theta -1)}{1536 r^{4}} + {\cal{O}}\left(\frac{1}{r^{5}}\right)\right]\,,
\\
\theta_{h}(r,\theta) &= \theta + {\rm cos}\theta \; {\rm sin}\theta \left[\frac{a^{2}}{2 r^{2}} + \frac{a^{2}}{2 r^{3}} + \frac{a^{4} (2 {\rm cos}^{2} \theta - 3)}{8 r^{4}} 
\right.
\nn \\
&\left.
+ \frac{201 \bar{\zeta} a^{2} m^{2}}{3584 r^{4}} + {\cal{O}}\left(\frac{1}{r^{5}}\right)\right]\,,
\end{align}
where $a = S/m$ and $\bar{\zeta} = \alpha^{2}/(\kappa \beta m^{4})$, with $\alpha = 4 \alpha_{4}$. Performing the coordinate transformation of the metric, we obtain
\begin{align}
\delta g_{00} &= 2 \delta U\,,
\\
\delta g_{0j} &= {\cal{O}}\left(\frac{1}{r_{h}^{5}}\right)\,,
\\
\delta g_{jk} &= \delta_{jk} \left(2 \delta U + 2 \delta V\right) + \delta W n_{j} n_{k} + \delta Y \hat{a}_{j} \hat{a}_{k}\,,
\end{align}
with $n_{j} = ({\rm sin}\theta \; {\rm cos}\phi, {\rm sin}\theta \; {\rm sin}\phi, {\rm cos}\theta)$ and $\hat{a}_{j} = (0, 0, 1)$. The explicit forms for the potentials are
\begin{align}
\delta U &= \frac{201}{3584} \bar{\zeta} \frac{a^{2} m}{r_{h}^{3}} (3 {\rm cos}^{2}\theta_{h} - 1) \left[1 - 2 \left(\frac{m}{r_{h}}\right) + {\cal{O}}\left(\frac{1}{r_{h}^{2}}\right)\right]\,,
\\
\delta V &= \frac{1}{2} \bar{\zeta} \frac{a^{2} m^{2}}{r_{h}^{4}} \left(\frac{1644}{1792} {\rm cos}^{2}\theta_{h} - \frac{11}{32}\right) + {\cal{O}}\left(\frac{1}{r_{h}^{5}}\right)\,,
\\
\delta W &= \bar{\zeta} \frac{a^{2} m^{2}}{r_{h}^{4}} \left(\frac{2295}{1792} {\rm cos}^{2}\theta_{h} - \frac{163}{256}\right) + {\cal{O}}\left(\frac{1}{r_{h}^{5}}\right)\,,
\\
\delta Y &= - \bar{\zeta} \frac{a^{2} m^{2}}{r_{h}^{4}} \left(\frac{389}{448} {\rm cos}^{2}\theta_{h} - \frac{201}{1792}\right) + {\cal{O}}\left(\frac{1}{r_{h}^{5}}\right)\,.
\end{align}
This completes the isolated BH metric in harmonic coordinates.

%%%%%%%%%%%%%%%%%%%%%%%%%%%%%%%%%%%%%%%%%%%%
%%%%%%%%%%%%%%%%%%%%%%%%%%%%%%%%%%%%%%%%%%%%
\bibliography{master}
\end{document}